\documentclass[prl,aps,twocolumn,preprintnumbers,amssymb,nobibnotes,nofootinbib,superscriptaddress,raggedbottom,10pt]{revtex4-1}
\usepackage{hyperref,ifthen,graphicx,array,multirow,color,amsmath,mathrsfs,titlesec,shuffle,tikz}
\usetikzlibrary{calc,decorations.markings,shapes.geometric}

\definecolor{mhvblue2}{rgb}{0.3,0.3,0.575}

\definecolor{mhvblue}{rgb}{0.6,0.6,0.7765}
\definecolor{nmhvred}{rgb}{0.6765,0.15,0.3}
\definecolor{ampgrey}{rgb}{0.9,0.9,0.9}
\definecolor{unord}{rgb}{0,0,0}
\definecolor{ord}{rgb}{0,0,0.575}
\definecolor{anchorLeg}{rgb}{0.575,0.0,0.225}
\definecolor{labelcolor}{rgb}{0,0,0}

\newcounter{legSteps}
\newcounter{offset}

\def\figScale{0.9}
\def\legSpread{4}
\def\extLegLen{0.9*0.32*\figScale}
\def\edgeLen{1*\figScale}

\pgfmathsetmacro{\pLen}{\edgeLen/(2*sin(72/2))}
\def\legLen{\edgeLen*0.45}

\def\labelDist{\legLen*1.5}
\def\lineThickness{(1pt)}
\def\dotSize{(\figScale*12pt)}
\def\ampSize{(1*\figScale*12pt)}
\def\eph{0.4}
\def\pageW{15.175/\figScale}

\tikzset{fullamp/.style={coordinate,minimum size=0.7*\ampSize,ball color=black!20,circle,text=white,inner sep=0}}
\tikzset{fullmhv/.style={coordinate,minimum size=0.8*\ampSize,ball color=mhvblue,circle,text=white,inner sep=0}}
\tikzset{fullmhvBig/.style={coordinate,minimum size=1*\ampSize,ball color=mhvblue,circle,text=white,inner sep=0}}
\tikzset{fullnmhv/.style={coordinate,minimum size=0.8*\ampSize,ball color=nmhvred,circle,text=white,inner sep=0}}
\tikzset{fullmhvBar/.style={coordinate,minimum size=0.8*\ampSize,ball color=white,circle,text=white,inner sep=0}}
\tikzset{ordAmp/.style={fill=ampgrey,circle,draw=black,line width=\lineThickness,minimum size=0.6*\ampSize,text=white,inner sep=0}}
\tikzset{mhv/.style={fill=mhvblue,circle,draw=black,line width=\lineThickness,minimum size=0.7*\ampSize,text=white,inner sep=0}}
\tikzset{nmhv/.style={fill=nmhvred,circle,draw=black,line width=\lineThickness,minimum size=0.7*\ampSize,text=white,inner sep=0}}
\tikzset{mhvBar/.style={fill=white,circle,draw=black,line width=\lineThickness,minimum size=0.7*\ampSize,text=white,inner sep=0}}
\tikzset{rEdge/.style={anchorLeg,line width=\lineThickness,line cap=round}}
\tikzset{fgraphEdge/.style={anchorLeg,line width=0.8*\lineThickness,line cap=round}}
\tikzset{fgraphExt/.style={ord,line width=\lineThickness,line cap=round}}
\tikzset{fgraphOpt/.style={ord,dotted,line width=\lineThickness,line cap=round}}
\tikzset{fdot/.style={fill=anchorLeg,circle,minimum size=0.35*\ampSize,inner sep=0}}
\tikzset{bdot/.style={fill=black,circle,minimum size=0.35*\ampSize,inner sep=0}}
\tikzset{ephdot/.style={fill=black,circle,minimum size=0.125*\ampSize,inner sep=0}}
\tikzset{ext/.style={black,line width=\lineThickness,line cap=round}}
\tikzset{under/.style={white,line width=4*\lineThickness,line cap=round}}
\tikzset{optExt/.style={black,dotted,line width=\lineThickness,line cap=round,rounded corners=10pt}}
\tikzset{optExtSc/.style={black,line width=\lineThickness,line cap=round,rounded corners=10pt}}
\tikzset{dashed/.style={black!70,dotted,line width=\lineThickness,line cap=round,rounded corners=10pt}}
\tikzset{ddot/.style={fill=black,circle,minimum size=0.35*\dotSize,inner sep=0}}
\tikzset{int/.style={black,line width=\lineThickness,line cap=round,rounded corners=1.5pt}}

\tikzset{intInfR/.style={nmhvred,line width=\lineThickness,line cap=round,rounded corners=1.5pt}}
\tikzset{blueDot/.style={fill=mhvblue,circle,draw=black,line width=\lineThickness,minimum size=0.5*\ampSize,text=white,inner sep=0}}
\tikzset{whiteDot/.style={fill=white,circle,draw=black,line width=\lineThickness,minimum size=0.5*\ampSize,text=white,inner sep=0}}
\tikzset{blackDot/.style={fill=black,circle,minimum size=0.5*\ampSize,inner sep=0}}
\tikzset{compositeDot/.style={fill=none,draw=black,line width=\lineThickness,circle,minimum size=0.75*\ampSize,inner sep=0}}

\tikzset{directedEdge/.style={draw=none,decoration={markings,mark connection node=connode,mark=at position 0.5 with {\node[transform shape, scale=0.205*\figScale,shape=dart,aspect=0.5,fill=black,draw] (connode) {};}},postaction={decorate}}}

\tikzset{directedEdgeBend/.style={rounded corners=10pt,draw=none,decoration={markings,mark connection node=connode,mark=at position 0.5 with {\node[transform shape, scale=0.205,shape=dart,aspect=0.5,fill=black,draw] (connode) {};}},postaction={decorate}}}

\newcommand{\leg}[3]{\draw[ext] #1--($#1+(#2:\legLen)$);\node at ($#1+(#2:\labelDist)$)[]{{\footnotesize #3}};}

\newcommand{\edgeA}{\text{{\footnotesize$a$}}}
\newcommand{\edgeB}{\text{{\footnotesize$b$}}}
\newcommand{\edgeC}{\text{{\footnotesize$c$}}}
\newcommand{\edgeD}{\text{{\footnotesize$d$}}}
\newcommand{\edgeE}{\text{{\footnotesize$e$}}}
\newcommand{\edgeF}{\text{{\footnotesize$f$}}}
\newcommand{\edgeG}{\text{{\footnotesize$g$}}}

\def\boundingDraw{red}
\def\boundingDraw{none}

\definecolor{legColour}{rgb}{0.35,0.35,0.35}
\definecolor{ndotColor}{rgb}{0.65,0.25,0.25}
\def\markStroke{0.65}

\tikzset{ndot/.style={transform shape,scale=0.35*\figScale,aspect=0.65,draw=ndotColor,line width=\markStroke*\figScale,shape=circle,fill=none}}

\tikzset{markedEdgeR/.style={draw=none,decoration={markings,mark connection node=connode,mark=at position 0.5 with {\node[ndot] (connode) {};}},postaction={decorate}}}

\tikzset{ndotR/.style={transform shape,scale=0.35*\figScale,aspect=0.65,draw=ndotColor,line width=\markStroke*\figScale,shape=circle,fill=white}}

\newcommand{\legAlt}[2]{
\fill[legColour] #1--($#1+(#2+\legSpread*3:\extLegLen)$)--($#1+(#2-\legSpread*3:\extLegLen)$);
\node at #1 [ddot]{};
}

\newcommand{\oneLoopGraphElement}[2][0]{\def\rotn{-360/#2}\def\edgeLen{0.75*\figScale}
\ifthenelse{#2=3}{\coordinate (v0) at (0,-\edgeLen/9)}{\coordinate (v0) at (0,0)};
\ifthenelse{#2=1}{\draw[int]($(v0)+(\edgeLen/2*0.3,0)$) arc (0:360:\edgeLen/2*0.3 and \edgeLen/1.125*0.3)coordinate[pos=0.25](e1);\legAlt{($(v0)-(0,\edgeLen/1.125*0.3)$)}{-90}}{\ifthenelse{#2=2}{
\draw[int]($(v0)+(\edgeLen/1.125*0.45,0)$) arc (0:360:\edgeLen/1.125*0.45 and \edgeLen/2*0.45)coordinate[pos=0.5](a1)coordinate[pos=0](a0)coordinate[pos=0.25](e1)coordinate[pos=0.75](e2);\legAlt{(a1)}{180};\legAlt{(a0)}{0};
}{
\foreach\a in {0,...,#2}{\coordinate (a\a) at ($(v0)+(\a*\rotn-90-\rotn/2:\edgeLen/2)$);};
\foreach\a[remember=\a as \la] in {0,...,#2}{\ifthenelse{\a=0}{}{\draw[int](a\la)--(a\a)coordinate (e\a) at ($(a\la)!0.5!(a\a)$);}};
\foreach\a in {1,...,#2}{\legAlt{(a\a)}{\a*\rotn-90-\rotn/2};};
}};
}

\newcommand{\oneLoopGraphElementLS}[2][0]{\def\extLegLen{1.1*0.32*\figScale}
\def\rotn{-360/#2}\def\edgeLen{0.75*\figScale}
\ifthenelse{#2=3}{\coordinate (v0) at (0,-\edgeLen/9)}{\coordinate (v0) at (0,0)};
\ifthenelse{#2=1}{\draw[int]($(v0)+(\edgeLen/2*0.3,0)$) arc (0:360:\edgeLen/2*0.3 and \edgeLen/1.125*0.3)coordinate[pos=0.25](e1);\legAlt{($(v0)-(0,\edgeLen/1.125*0.3)$)}{-90}}{\ifthenelse{#2=2}{
\draw[int]($(v0)+(\edgeLen/1.125*0.45,0)$) arc (0:360:\edgeLen/1.125*0.45 and \edgeLen/2*0.45)coordinate[pos=0.5](a1)coordinate[pos=0](a0)coordinate[pos=0.25](e1)coordinate[pos=0.75](e2);\legAlt{(a1)}{180};\legAlt{(a0)}{0};
}{
\foreach\a in {0,...,#2}{\coordinate (a\a) at ($(v0)+(\a*\rotn-90-\rotn/2:\edgeLen/2)$);};
\foreach\a[remember=\a as \la] in {0,...,#2}{\ifthenelse{\a=0}{}{\draw[int](a\la)--(a\a)coordinate (e\a) at ($(a\la)!0.5!(a\a)$);}};
\foreach\a in {1,...,#2}{\legAlt{(a\a)}{\a*\rotn-90-\rotn/2};};
}};\def\extLegLen{0.9*0.32*\figScale}
}

\newcommand{\intDots}[1]{
\foreach\n in {1,...,#1}{\node at (v\n) [bdot] {};};}%\node at (v2) [bdot] {};\node at (v3) [bdot] {};\node at (v4) [bdot] {};\node at (v5) [bdot] {};\node at (v6) [bdot] {};\node at (v7) [bdot] {};}

\newcommand{\lsVerts}[7]{
\ifthenelse{#1=1}{\node at (v1) [fullmhvBar] {};}{\ifthenelse{#1=2}{\node at (v1) [fullmhv] {};}{\ifthenelse{#1=3}{\node at (v1) [fullnmhv] {};}{\ifthenelse{#1=0}{\node at (v1) [bdot] {};}{}}}}
\ifthenelse{#2=1}{\node at (v2) [fullmhvBar] {};}{\ifthenelse{#2=2}{\node at (v2) [fullmhv] {};}{\ifthenelse{#2=3}{\node at (v2) [fullnmhv] {};}{\ifthenelse{#2=0}{\node at (v2) [bdot] {};}{}}}}
\ifthenelse{#3=1}{\node at (v3) [fullmhvBar] {};}{\ifthenelse{#3=2}{\node at (v3) [fullmhv] {};}{\ifthenelse{#3=3}{\node at (v3) [fullnmhv] {};}{\ifthenelse{#3=0}{\node at (v3) [bdot] {};}{}}}}
\ifthenelse{#4=1}{\node at (v4) [fullmhvBar] {};}{\ifthenelse{#4=2}{\node at (v4) [fullmhv] {};}{\ifthenelse{#4=3}{\node at (v4) [fullnmhv] {};}{\ifthenelse{#4=0}{\node at (v4) [bdot] {};}{}}}}
\ifthenelse{#5=1}{\node at (v5) [fullmhvBar] {};}{\ifthenelse{#5=2}{\node at (v5) [fullmhv] {};}{\ifthenelse{#5=3}{\node at (v5) [fullnmhv] {};}{\ifthenelse{#5=0}{\node at (v5) [bdot] {};}{}}}}
\ifthenelse{#6=1}{\node at (v6) [fullmhvBar] {};}{\ifthenelse{#6=2}{\node at (v6) [fullmhv] {};}{\ifthenelse{#6=3}{\node at (v6) [fullnmhv] {};}{\ifthenelse{#6=0}{\node at (v6) [bdot] {};}{}}}}
\ifthenelse{#7=1}{\node at (v7) [fullmhvBar] {};}{\ifthenelse{#7=2}{\node at (v7) [fullmhv] {};}{\ifthenelse{#7=3}{\node at (v7) [fullnmhv] {};}{\ifthenelse{#7=0}{\node at (v7) [bdot] {};}{}}}}}

\newcommand{\lsVertsOrd}[7]{
\ifthenelse{#1=1}{\node at (v1) [mhvBar] {};}{\ifthenelse{#1=2}{\node at (v1) [mhv] {};}{\ifthenelse{#1=3}{\node at (v1) [nmhv] {};}{\ifthenelse{#1=0}{\node at (v1) [bdot] {};}{}}}}
\ifthenelse{#2=1}{\node at (v2) [mhvBar] {};}{\ifthenelse{#2=2}{\node at (v2) [mhv] {};}{\ifthenelse{#2=3}{\node at (v2) [nmhv] {};}{\ifthenelse{#2=0}{\node at (v2) [bdot] {};}{}}}}
\ifthenelse{#3=1}{\node at (v3) [mhvBar] {};}{\ifthenelse{#3=2}{\node at (v3) [mhv] {};}{\ifthenelse{#3=3}{\node at (v3) [nmhv] {};}{\ifthenelse{#3=0}{\node at (v3) [bdot] {};}{}}}}
\ifthenelse{#4=1}{\node at (v4) [mhvBar] {};}{\ifthenelse{#4=2}{\node at (v4) [mhv] {};}{\ifthenelse{#4=3}{\node at (v4) [nmhv] {};}{\ifthenelse{#4=0}{\node at (v4) [bdot] {};}{}}}}
\ifthenelse{#5=1}{\node at (v5) [mhvBar] {};}{\ifthenelse{#5=2}{\node at (v5) [mhv] {};}{\ifthenelse{#5=3}{\node at (v5) [nmhv] {};}{\ifthenelse{#5=0}{\node at (v5) [bdot] {};}{}}}}
\ifthenelse{#6=1}{\node at (v6) [mhvBar] {};}{\ifthenelse{#6=2}{\node at (v6) [mhv] {};}{\ifthenelse{#6=3}{\node at (v6) [nmhv] {};}{\ifthenelse{#6=0}{\node at (v6) [bdot] {};}{}}}}
\ifthenelse{#7=1}{\node at (v7) [mhvBar] {};}{\ifthenelse{#7=2}{\node at (v7) [mhv] {};}{\ifthenelse{#7=3}{\node at (v7) [fullnmhv] {};}{\ifthenelse{#7=0}{\node at (v7) [bdot] {};}{}}}}}

\newcommand{\colorF}[1]{\begin{tikzpicture}[scale=\figScale,baseline=-3.05]\useasboundingbox ($(-1.4,-1.4)$) rectangle ($(1.4,1.4)$);\draw[int,line width=0.1,red,draw=\boundingDraw] ($(-1.4,-1.4)$) rectangle ($(1.4,1.4)$);\draw[fgraphEdge] (0,0) circle (0.7);\draw[fgraphEdge] (-90:0.7)--(90:0.7);\node at (-90:0.7) [fdot] {};\node at (90:0.7) [fdot] {};
#1\end{tikzpicture}}

\newcommand{\dPent}[1]{\begin{tikzpicture}[scale=\figScale,baseline=-3.05]\useasboundingbox ($(-\pageW/9,-1.4)$) rectangle ($(\pageW/9,1.4)$);\draw[int,line width=0.1,red,draw=\boundingDraw] ($(-\pageW/9,-1.4)$) rectangle ($(\pageW/9,1.4)$);\coordinate(v7)at($\figScale*(0,0)$);\coordinate(v6)at($(v7)+(-90:\figScale*0.65)$);\coordinate(v3)at($(v7)+(90:\figScale*0.65)$);\coordinate(v2)at($(v3)+(180:\figScale*1.25)$);\coordinate(v4)at($(v3)+(0:\figScale*1.25)$);\coordinate(v1)at($(v6)+(180:\figScale*1.25)$);\coordinate(v5)at($(v6)+(0:\figScale*1.25)$);
#1\end{tikzpicture}}

\newcommand{\npPbox}[1]{\begin{tikzpicture}[scale=\figScale,baseline=-3.05]\useasboundingbox ($(-\pageW/9,-1.5)$) rectangle ($(\pageW/9,1.5)$);\draw[int,line width=0.1,red,draw=\boundingDraw] ($(-\pageW/9,-1.5)$) rectangle ($(\pageW/9,1.5)$);\coordinate(v7)at($\figScale*(-1.85,0)$);\coordinate(v1)at($(v7)+(-54:\figScale*1.119)$);\coordinate(v2)at($(v7)+(54:\figScale*1.119)$);\coordinate(v3)at($\figScale*(0.35,0.65)$);\coordinate(v4)at($\figScale*(1.0,-0)$);\coordinate(v5)at($\figScale*(0.35,-0.65)$);\coordinate(v6)at($\figScale*(-0.2,0)$);
#1\end{tikzpicture}}

\newcommand{\dPentPlainEdges}{\draw[int](v6)--(v1);\draw[int](v1)--(v2);\draw[int](v2)--(v3);\draw[int](v3)--(v4);\draw[int](v4)--(v5);\draw[int](v5)--(v6);\draw[int](v6)--(v7);\draw[int](v7)--(v3);}

\newcommand{\npPboxPlainEdges}{\draw[int](v5)--(v1);\draw[int](v1)--(v2);\draw[int](v2)--(v3);\draw[int](v3)--(v4);\draw[int](v4)--(v5);\draw[int](v5)--(v6);\draw[int](v6)--(v3);}
\newcommand{\npPboxScalarEdges}{\npPboxPlainEdges
\draw[directedEdge](v5)--(v1);\node[inner sep=1pt,anchor=north west] at (connode) [] {\edgeA};\draw[directedEdge](v1)--(v2);\node[inner sep=2.5pt,anchor=east] at (connode) [] {\edgeB};\draw[directedEdge](v2)--(v3);\node[inner sep=1pt,anchor=south west] at (connode) [] {\edgeC};\draw[directedEdge](v3)--(v4);\node[inner sep=1pt,anchor=south west] at (connode) [] {\edgeD};\draw[directedEdge](v4)--(v5);\node[inner sep=1pt,anchor=north west] at (connode) [] {\edgeE};\draw[directedEdge](v5)--(v6);\node[inner sep=1pt,anchor=east] at (connode) [] {\edgeF\;};\draw[directedEdge](v6)--(v3);\node[inner sep=1pt,anchor=east] at (connode) [] {\edgeG\;};}

\newcommand{\kBoxLegs}[7]{
\setcounter{legSteps}{0}
\def\zeroAngle{-90}\def\spread{45}\setcounter{offset}{-1}\addtocounter{offset}{#1}
\ifthenelse{#1=0}{}{\ifthenelse{#1=1}{\stepcounter{legSteps}\leg{(v1)}{\zeroAngle}{\arabic{legSteps}};}{
\foreach\n in {1,...,#1}{\def\eph{\arabic{offset}}\def\angle{\zeroAngle-2*\n*\spread/\eph+#1*\spread/\eph+\spread/\eph}\stepcounter{legSteps}\leg{(v1)}{\angle}{\arabic{legSteps}}}}}

\def\zeroAngle{180}\def\spread{45}\setcounter{offset}{-1}\addtocounter{offset}{#2}
\ifthenelse{#2=0}{}{\ifthenelse{#2=1}{\stepcounter{legSteps}\leg{(v2)}{\zeroAngle}{\arabic{legSteps}};}{
\foreach\n in {1,...,#2}{\def\eph{\arabic{offset}}\def\angle{\zeroAngle-2*\n*\spread/\eph+#2*\spread/\eph+\spread/\eph}\stepcounter{legSteps}\leg{(v2)}{\angle}{\arabic{legSteps}}}}}

\def\zeroAngle{90}\def\spread{45}\setcounter{offset}{-1}\addtocounter{offset}{#3}
\ifthenelse{#3=0}{}{\ifthenelse{#3=1}{\stepcounter{legSteps}\leg{(v3)}{\zeroAngle}{\arabic{legSteps}};}{
\foreach\n in {1,...,#3}{\def\eph{\arabic{offset}}\def\angle{\zeroAngle-2*\n*\spread/\eph+#3*\spread/\eph+\spread/\eph}\stepcounter{legSteps}\leg{(v3)}{\angle}{\arabic{legSteps}}}}}

\def\zeroAngle{90}\def\spread{15}\setcounter{offset}{-1}\addtocounter{offset}{#4}
\ifthenelse{#4=0}{}{\ifthenelse{#4=1}{\stepcounter{legSteps}\leg{(v4)}{\zeroAngle}{\arabic{legSteps}};}{
\foreach\n in {1,...,#4}{\def\eph{\arabic{offset}}\def\angle{\zeroAngle-2*\n*\spread/\eph+#4*\spread/\eph+\spread/\eph}\stepcounter{legSteps}\leg{(v4)}{\angle}{\arabic{legSteps}}}}}

\def\zeroAngle{90}\def\spread{45}\setcounter{offset}{-1}\addtocounter{offset}{#5}
\ifthenelse{#5=0}{}{\ifthenelse{#5=1}{\stepcounter{legSteps}\leg{(v5)}{\zeroAngle}{\arabic{legSteps}};}{
\foreach\n in {1,...,#5}{\def\eph{\arabic{offset}}\def\angle{\zeroAngle-2*\n*\spread/\eph+#5*\spread/\eph+\spread/\eph}\stepcounter{legSteps}\leg{(v5)}{\angle}{\arabic{legSteps}}}}}

\def\zeroAngle{0}\def\spread{45}\setcounter{offset}{-1}\addtocounter{offset}{#6}
\ifthenelse{#6=0}{}{\ifthenelse{#6=1}{\stepcounter{legSteps}\leg{(v6)}{\zeroAngle}{\arabic{legSteps}};}{
\foreach\n in {1,...,#6}{\def\eph{\arabic{offset}}\def\angle{\zeroAngle-2*\n*\spread/\eph+#6*\spread/\eph+\spread/\eph}\stepcounter{legSteps}\leg{(v6)}{\angle}{\arabic{legSteps}}}}}

\def\zeroAngle{-90}\def\spread{45}\setcounter{offset}{-1}\addtocounter{offset}{#7}
\ifthenelse{#7=0}{}{\ifthenelse{#7=1}{\stepcounter{legSteps}\leg{(v7)}{\zeroAngle}{\arabic{legSteps}};}{
\foreach\n in {1,...,#7}{\def\eph{\arabic{offset}}\def\angle{\zeroAngle-2*\n*\spread/\eph+#7*\spread/\eph+\spread/\eph}\stepcounter{legSteps}\leg{(v7)}{\angle}{\arabic{legSteps}}}}}
}

\newcommand{\kTboxLegs}[6]{
\setcounter{legSteps}{0}
\def\zeroAngle{-135}\def\spread{45}\setcounter{offset}{-1}\addtocounter{offset}{#1}
\ifthenelse{#1=0}{}{\ifthenelse{#1=1}{\stepcounter{legSteps}\leg{(v1)}{\zeroAngle}{\arabic{legSteps}};}{
\foreach\n in {1,...,#1}{\def\eph{\arabic{offset}}\def\angle{\zeroAngle-2*\n*\spread/\eph+#1*\spread/\eph+\spread/\eph}\stepcounter{legSteps}\leg{(v1)}{\angle}{\arabic{legSteps}}}}}

\def\zeroAngle{135}\def\spread{45}\setcounter{offset}{-1}\addtocounter{offset}{#2}
\ifthenelse{#2=0}{}{\ifthenelse{#2=1}{\stepcounter{legSteps}\leg{(v2)}{\zeroAngle}{\arabic{legSteps}};}{
\foreach\n in {1,...,#2}{\def\eph{\arabic{offset}}\def\angle{\zeroAngle-2*\n*\spread/\eph+#2*\spread/\eph+\spread/\eph}\stepcounter{legSteps}\leg{(v2)}{\angle}{\arabic{legSteps}}}}}

\def\zeroAngle{90}\def\spread{15}\setcounter{offset}{-1}\addtocounter{offset}{#3}
\ifthenelse{#3=0}{}{\ifthenelse{#3=1}{\stepcounter{legSteps}\leg{(v3)}{\zeroAngle}{\arabic{legSteps}};}{
\foreach\n in {1,...,#3}{\def\eph{\arabic{offset}}\def\angle{\zeroAngle-2*\n*\spread/\eph+#3*\spread/\eph+\spread/\eph}\stepcounter{legSteps}\leg{(v3)}{\angle}{\arabic{legSteps}}}}}

\def\zeroAngle{90}\def\spread{45}\setcounter{offset}{-1}\addtocounter{offset}{#4}
\ifthenelse{#4=0}{}{\ifthenelse{#4=1}{\stepcounter{legSteps}\leg{(v4)}{\zeroAngle}{\arabic{legSteps}};}{
\foreach\n in {1,...,#4}{\def\eph{\arabic{offset}}\def\angle{\zeroAngle-2*\n*\spread/\eph+#4*\spread/\eph+\spread/\eph}\stepcounter{legSteps}\leg{(v4)}{\angle}{\arabic{legSteps}}}}}

\def\zeroAngle{0}\def\spread{45}\setcounter{offset}{-1}\addtocounter{offset}{#5}
\ifthenelse{#5=0}{}{\ifthenelse{#5=1}{\stepcounter{legSteps}\leg{(v5)}{\zeroAngle}{\arabic{legSteps}};}{
\foreach\n in {1,...,#5}{\def\eph{\arabic{offset}}\def\angle{\zeroAngle-2*\n*\spread/\eph+#5*\spread/\eph+\spread/\eph}\stepcounter{legSteps}\leg{(v5)}{\angle}{\arabic{legSteps}}}}}

\def\zeroAngle{-90}\def\spread{45}\setcounter{offset}{-1}\addtocounter{offset}{#6}
\ifthenelse{#6=0}{}{\ifthenelse{#6=1}{\stepcounter{legSteps}\leg{(v6)}{\zeroAngle}{\arabic{legSteps}};}{
\foreach\n in {1,...,#6}{\def\eph{\arabic{offset}}\def\angle{\zeroAngle-2*\n*\spread/\eph+#6*\spread/\eph+\spread/\eph}\stepcounter{legSteps}\leg{(v6)}{\angle}{\arabic{legSteps}}}}}
}

\newcommand{\kTLegs}[5]{
\setcounter{legSteps}{0}
\def\zeroAngle{-135}\def\spread{45}\setcounter{offset}{-1}\addtocounter{offset}{#1}
\ifthenelse{#1=0}{}{\ifthenelse{#1=1}{\stepcounter{legSteps}\leg{(v1)}{\zeroAngle}{\arabic{legSteps}};}{
\foreach\n in {1,...,#1}{\def\eph{\arabic{offset}}\def\angle{\zeroAngle-2*\n*\spread/\eph+#1*\spread/\eph+\spread/\eph}\stepcounter{legSteps}\leg{(v1)}{\angle}{\arabic{legSteps}}}}}

\def\zeroAngle{135}\def\spread{45}\setcounter{offset}{-1}\addtocounter{offset}{#2}
\ifthenelse{#2=0}{}{\ifthenelse{#2=1}{\stepcounter{legSteps}\leg{(v2)}{\zeroAngle}{\arabic{legSteps}};}{
\foreach\n in {1,...,#2}{\def\eph{\arabic{offset}}\def\angle{\zeroAngle-2*\n*\spread/\eph+#2*\spread/\eph+\spread/\eph}\stepcounter{legSteps}\leg{(v2)}{\angle}{\arabic{legSteps}}}}}

\def\zeroAngle{90}\def\spread{15}\setcounter{offset}{-1}\addtocounter{offset}{#3}
\ifthenelse{#3=0}{}{\ifthenelse{#3=1}{\stepcounter{legSteps}\leg{(v3)}{\zeroAngle}{\arabic{legSteps}};}{
\foreach\n in {1,...,#3}{\def\eph{\arabic{offset}}\def\angle{\zeroAngle-2*\n*\spread/\eph+#3*\spread/\eph+\spread/\eph}\stepcounter{legSteps}\leg{(v3)}{\angle}{\arabic{legSteps}}}}}

\def\zeroAngle{45}\def\spread{45}\setcounter{offset}{-1}\addtocounter{offset}{#4}
\ifthenelse{#4=0}{}{\ifthenelse{#4=1}{\stepcounter{legSteps}\leg{(v4)}{\zeroAngle}{\arabic{legSteps}};}{
\foreach\n in {1,...,#4}{\def\eph{\arabic{offset}}\def\angle{\zeroAngle-2*\n*\spread/\eph+#4*\spread/\eph+\spread/\eph}\stepcounter{legSteps}\leg{(v4)}{\angle}{\arabic{legSteps}}}}}

\def\zeroAngle{-45}\def\spread{45}\setcounter{offset}{-1}\addtocounter{offset}{#5}
\ifthenelse{#5=0}{}{\ifthenelse{#5=1}{\stepcounter{legSteps}\leg{(v5)}{\zeroAngle}{\arabic{legSteps}};}{
\foreach\n in {1,...,#5}{\def\eph{\arabic{offset}}\def\angle{\zeroAngle-2*\n*\spread/\eph+#5*\spread/\eph+\spread/\eph}\stepcounter{legSteps}\leg{(v5)}{\angle}{\arabic{legSteps}}}}}
}

\newcommand{\pBoxLegs}[7]{
\setcounter{legSteps}{0}
\def\zeroAngle{-108}\def\spread{45}\setcounter{offset}{-1}\addtocounter{offset}{#1}
\ifthenelse{#1=0}{}{\ifthenelse{#1=1}{\stepcounter{legSteps}\leg{(v1)}{\zeroAngle}{\arabic{legSteps}};}{
\foreach\n in {1,...,#1}{\def\eph{\arabic{offset}}\def\angle{\zeroAngle-2*\n*\spread/\eph+#1*\spread/\eph+\spread/\eph}\stepcounter{legSteps}\leg{(v1)}{\angle}{\arabic{legSteps}}}}}

\def\zeroAngle{180}\def\spread{45}\setcounter{offset}{-1}\addtocounter{offset}{#2}
\ifthenelse{#2=0}{}{\ifthenelse{#2=1}{\stepcounter{legSteps}\leg{(v2)}{\zeroAngle}{\arabic{legSteps}};}{
\foreach\n in {1,...,#2}{\def\eph{\arabic{offset}}\def\angle{\zeroAngle-2*\n*\spread/\eph+#2*\spread/\eph+\spread/\eph}\stepcounter{legSteps}\leg{(v2)}{\angle}{\arabic{legSteps}}}}}

\def\zeroAngle{108}\def\spread{45}\setcounter{offset}{-1}\addtocounter{offset}{#3}
\ifthenelse{#3=0}{}{\ifthenelse{#3=1}{\stepcounter{legSteps}\leg{(v3)}{\zeroAngle}{\arabic{legSteps}};}{
\foreach\n in {1,...,#3}{\def\eph{\arabic{offset}}\def\angle{\zeroAngle-2*\n*\spread/\eph+#3*\spread/\eph+\spread/\eph}\stepcounter{legSteps}\leg{(v3)}{\angle}{\arabic{legSteps}}}}}

\def\zeroAngle{81}\def\spread{20}\setcounter{offset}{-1}\addtocounter{offset}{#4}
\ifthenelse{#4=0}{}{\ifthenelse{#4=1}{\stepcounter{legSteps}\leg{(v4)}{\zeroAngle}{\arabic{legSteps}};}{
\foreach\n in {1,...,#4}{\def\eph{\arabic{offset}}\def\angle{\zeroAngle-2*\n*\spread/\eph+#4*\spread/\eph+\spread/\eph}\stepcounter{legSteps}\leg{(v4)}{\angle}{\arabic{legSteps}}}}}

\def\zeroAngle{45}\def\spread{45}\setcounter{offset}{-1}\addtocounter{offset}{#5}
\ifthenelse{#5=0}{}{\ifthenelse{#5=1}{\stepcounter{legSteps}\leg{(v5)}{\zeroAngle}{\arabic{legSteps}};}{
\foreach\n in {1,...,#5}{\def\eph{\arabic{offset}}\def\angle{\zeroAngle-2*\n*\spread/\eph+#5*\spread/\eph+\spread/\eph}\stepcounter{legSteps}\leg{(v5)}{\angle}{\arabic{legSteps}}}}}

\def\zeroAngle{-45}\def\spread{45}\setcounter{offset}{-1}\addtocounter{offset}{#6}
\ifthenelse{#6=0}{}{\ifthenelse{#6=1}{\stepcounter{legSteps}\leg{(v6)}{\zeroAngle}{\arabic{legSteps}};}{
\foreach\n in {1,...,#6}{\def\eph{\arabic{offset}}\def\angle{\zeroAngle-2*\n*\spread/\eph+#6*\spread/\eph+\spread/\eph}\stepcounter{legSteps}\leg{(v6)}{\angle}{\arabic{legSteps}}}}}

\def\zeroAngle{-81}\def\spread{20}\setcounter{offset}{-1}\addtocounter{offset}{#7}
\ifthenelse{#7=0}{}{\ifthenelse{#7=1}{\stepcounter{legSteps}\leg{(v7)}{\zeroAngle}{\arabic{legSteps}};}{
\foreach\n in {1,...,#7}{\def\eph{\arabic{offset}}\def\angle{\zeroAngle-2*\n*\spread/\eph+#7*\spread/\eph+\spread/\eph}\stepcounter{legSteps}\leg{(v7)}{\angle}{\arabic{legSteps}}}}}
}

\newcommand{\hBoxLegs}[7]{
\setcounter{legSteps}{0}
\def\zeroAngle{-115}\def\spread{45}\setcounter{offset}{-1}\addtocounter{offset}{#1}
\ifthenelse{#1=0}{}{\ifthenelse{#1=1}{\stepcounter{legSteps}\leg{(v1)}{\zeroAngle}{\arabic{legSteps}};}{
\foreach\n in {1,...,#1}{\def\eph{\arabic{offset}}\def\angle{\zeroAngle-2*\n*\spread/\eph+#1*\spread/\eph+\spread/\eph}\stepcounter{legSteps}\leg{(v1)}{\angle}{\arabic{legSteps}}}}}

\def\zeroAngle{180}\def\spread{45}\setcounter{offset}{-1}\addtocounter{offset}{#2}
\ifthenelse{#2=0}{}{\ifthenelse{#2=1}{\stepcounter{legSteps}\leg{(v2)}{\zeroAngle}{\arabic{legSteps}};}{
\foreach\n in {1,...,#2}{\def\eph{\arabic{offset}}\def\angle{\zeroAngle-2*\n*\spread/\eph+#2*\spread/\eph+\spread/\eph}\stepcounter{legSteps}\leg{(v2)}{\angle}{\arabic{legSteps}}}}}

\def\zeroAngle{115}\def\spread{45}\setcounter{offset}{-1}\addtocounter{offset}{#3}
\ifthenelse{#3=0}{}{\ifthenelse{#3=1}{\stepcounter{legSteps}\leg{(v3)}{\zeroAngle}{\arabic{legSteps}};}{
\foreach\n in {1,...,#3}{\def\eph{\arabic{offset}}\def\angle{\zeroAngle-2*\n*\spread/\eph+#3*\spread/\eph+\spread/\eph}\stepcounter{legSteps}\leg{(v3)}{\angle}{\arabic{legSteps}}}}}

\def\zeroAngle{61.5}\def\spread{30}\setcounter{offset}{-1}\addtocounter{offset}{#4}
\ifthenelse{#4=0}{}{\ifthenelse{#4=1}{\stepcounter{legSteps}\leg{(v4)}{\zeroAngle}{\arabic{legSteps}};}{
\foreach\n in {1,...,#4}{\def\eph{\arabic{offset}}\def\angle{\zeroAngle-2*\n*\spread/\eph+#4*\spread/\eph+\spread/\eph}\stepcounter{legSteps}\leg{(v4)}{\angle}{\arabic{legSteps}}}}}

\def\zeroAngle{0}\def\spread{45}\setcounter{offset}{-1}\addtocounter{offset}{#5}
\ifthenelse{#5=0}{}{\ifthenelse{#5=1}{\stepcounter{legSteps}\leg{(v5)}{\zeroAngle}{\arabic{legSteps}};}{
\foreach\n in {1,...,#5}{\def\eph{\arabic{offset}}\def\angle{\zeroAngle-2*\n*\spread/\eph+#5*\spread/\eph+\spread/\eph}\stepcounter{legSteps}\leg{(v5)}{\angle}{\arabic{legSteps}}}}}

\def\zeroAngle{-61.5}\def\spread{30}\setcounter{offset}{-1}\addtocounter{offset}{#6}
\ifthenelse{#6=0}{}{\ifthenelse{#6=1}{\stepcounter{legSteps}\leg{(v6)}{\zeroAngle}{\arabic{legSteps}};}{
\foreach\n in {1,...,#6}{\def\eph{\arabic{offset}}\def\angle{\zeroAngle-2*\n*\spread/\eph+#6*\spread/\eph+\spread/\eph}\stepcounter{legSteps}\leg{(v6)}{\angle}{\arabic{legSteps}}}}}

\def\zeroAngle{180}\def\spread{25}\setcounter{offset}{-1}\addtocounter{offset}{#7}
\ifthenelse{#7=0}{}{\ifthenelse{#7=1}{\stepcounter{legSteps}\leg{(v7)}{\zeroAngle}{\arabic{legSteps}};}{
\foreach\n in {1,...,#7}{\def\eph{\arabic{offset}}\def\angle{\zeroAngle-2*\n*\spread/\eph+#7*\spread/\eph+\spread/\eph}\stepcounter{legSteps}\leg{(v7)}{\angle}{\arabic{legSteps}}}}}
}

\newcommand{\npPboxLegs}[6]{
\setcounter{legSteps}{0}
\def\zeroAngle{-135}\def\spread{45}\setcounter{offset}{-1}\addtocounter{offset}{#1}
\ifthenelse{#1=0}{}{\ifthenelse{#1=1}{\stepcounter{legSteps}\leg{(v1)}{\zeroAngle}{{\color{labelcolor}\arabic{legSteps}}};}{
\foreach\n in {1,...,#1}{\def\eph{\arabic{offset}}\def\angle{\zeroAngle-2*\n*\spread/\eph+#1*\spread/\eph+\spread/\eph}\stepcounter{legSteps}\leg{(v1)}{\angle}{{\color{labelcolor}\arabic{legSteps}}}}}}

\def\zeroAngle{135}\def\spread{45}\setcounter{offset}{-1}\addtocounter{offset}{#2}
\ifthenelse{#2=0}{}{\ifthenelse{#2=1}{\stepcounter{legSteps}\leg{(v2)}{\zeroAngle}{{\color{labelcolor}\arabic{legSteps}}};}{
\foreach\n in {1,...,#2}{\def\eph{\arabic{offset}}\def\angle{\zeroAngle-2*\n*\spread/\eph+#2*\spread/\eph+\spread/\eph}\stepcounter{legSteps}\leg{(v2)}{\angle}{{\color{labelcolor}\arabic{legSteps}}}}}}

\def\zeroAngle{61.5}\def\spread{30}\setcounter{offset}{-1}\addtocounter{offset}{#3}
\ifthenelse{#3=0}{}{\ifthenelse{#3=1}{\stepcounter{legSteps}\leg{(v3)}{\zeroAngle}{{\color{labelcolor}\arabic{legSteps}}};}{
\foreach\n in {1,...,#3}{\def\eph{\arabic{offset}}\def\angle{\zeroAngle-2*\n*\spread/\eph+#3*\spread/\eph+\spread/\eph}\stepcounter{legSteps}\leg{(v3)}{\angle}{{\color{labelcolor}\arabic{legSteps}}}}}}

\def\zeroAngle{0}\def\spread{45}\setcounter{offset}{-1}\addtocounter{offset}{#4}
\ifthenelse{#4=0}{}{\ifthenelse{#4=1}{\stepcounter{legSteps}\leg{(v4)}{\zeroAngle}{{\color{labelcolor}\arabic{legSteps}}};}{
\foreach\n in {1,...,#4}{\def\eph{\arabic{offset}}\def\angle{\zeroAngle-2*\n*\spread/\eph+#4*\spread/\eph+\spread/\eph}\stepcounter{legSteps}\leg{(v4)}{\angle}{{\color{labelcolor}\arabic{legSteps}}}}}}

\def\zeroAngle{-61.5}\def\spread{30}\setcounter{offset}{-1}\addtocounter{offset}{#5}
\ifthenelse{#5=0}{}{\ifthenelse{#5=1}{\stepcounter{legSteps}\leg{(v5)}{\zeroAngle}{{\color{labelcolor}\arabic{legSteps}}};}{
\foreach\n in {1,...,#5}{\def\eph{\arabic{offset}}\def\angle{\zeroAngle-2*\n*\spread/\eph+#5*\spread/\eph+\spread/\eph}\stepcounter{legSteps}\leg{(v5)}{\angle}{{\color{labelcolor}\arabic{legSteps}}}}}}

\def\zeroAngle{180}\def\spread{25}\setcounter{offset}{-1}\addtocounter{offset}{#6}
\ifthenelse{#6=0}{}{\ifthenelse{#6=1}{\stepcounter{legSteps}\leg{(v6)}{\zeroAngle}{{\color{labelcolor}\arabic{legSteps}}};}{
\foreach\n in {1,...,#6}{\def\eph{\arabic{offset}}\def\angle{\zeroAngle-2*\n*\spread/\eph+#6*\spread/\eph+\spread/\eph}\stepcounter{legSteps}\leg{(v6)}{\angle}{{\color{labelcolor}\arabic{legSteps}}}}}}
}

\newcommand{\pTLegs}[6]{
\setcounter{legSteps}{0}
\def\zeroAngle{-115}\def\spread{45}\setcounter{offset}{-1}\addtocounter{offset}{#1}
\ifthenelse{#1=0}{}{\ifthenelse{#1=1}{\stepcounter{legSteps}\leg{(v1)}{\zeroAngle}{\arabic{legSteps}};}{
\foreach\n in {1,...,#1}{\def\eph{\arabic{offset}}\def\angle{\zeroAngle-2*\n*\spread/\eph+#1*\spread/\eph+\spread/\eph}\stepcounter{legSteps}\leg{(v1)}{\angle}{\arabic{legSteps}}}}}

\def\zeroAngle{180}\def\spread{45}\setcounter{offset}{-1}\addtocounter{offset}{#2}
\ifthenelse{#2=0}{}{\ifthenelse{#2=1}{\stepcounter{legSteps}\leg{(v2)}{\zeroAngle}{\arabic{legSteps}};}{
\foreach\n in {1,...,#2}{\def\eph{\arabic{offset}}\def\angle{\zeroAngle-2*\n*\spread/\eph+#2*\spread/\eph+\spread/\eph}\stepcounter{legSteps}\leg{(v2)}{\angle}{\arabic{legSteps}}}}}

\def\zeroAngle{115}\def\spread{45}\setcounter{offset}{-1}\addtocounter{offset}{#3}
\ifthenelse{#3=0}{}{\ifthenelse{#3=1}{\stepcounter{legSteps}\leg{(v3)}{\zeroAngle}{\arabic{legSteps}};}{
\foreach\n in {1,...,#3}{\def\eph{\arabic{offset}}\def\angle{\zeroAngle-2*\n*\spread/\eph+#3*\spread/\eph+\spread/\eph}\stepcounter{legSteps}\leg{(v3)}{\angle}{\arabic{legSteps}}}}}

\def\zeroAngle{61.5}\def\spread{30}\setcounter{offset}{-1}\addtocounter{offset}{#4}
\ifthenelse{#4=0}{}{\ifthenelse{#4=1}{\stepcounter{legSteps}\leg{(v4)}{\zeroAngle}{\arabic{legSteps}};}{
\foreach\n in {1,...,#4}{\def\eph{\arabic{offset}}\def\angle{\zeroAngle-2*\n*\spread/\eph+#4*\spread/\eph+\spread/\eph}\stepcounter{legSteps}\leg{(v4)}{\angle}{\arabic{legSteps}}}}}

\def\zeroAngle{0}\def\spread{45}\setcounter{offset}{-1}\addtocounter{offset}{#5}
\ifthenelse{#5=0}{}{\ifthenelse{#5=1}{\stepcounter{legSteps}\leg{(v5)}{\zeroAngle}{\arabic{legSteps}};}{
\foreach\n in {1,...,#5}{\def\eph{\arabic{offset}}\def\angle{\zeroAngle-2*\n*\spread/\eph+#5*\spread/\eph+\spread/\eph}\stepcounter{legSteps}\leg{(v5)}{\angle}{\arabic{legSteps}}}}}

\def\zeroAngle{-61.5}\def\spread{30}\setcounter{offset}{-1}\addtocounter{offset}{#6}
\ifthenelse{#6=0}{}{\ifthenelse{#6=1}{\stepcounter{legSteps}\leg{(v6)}{\zeroAngle}{\arabic{legSteps}};}{
\foreach\n in {1,...,#6}{\def\eph{\arabic{offset}}\def\angle{\zeroAngle-2*\n*\spread/\eph+#6*\spread/\eph+\spread/\eph}\stepcounter{legSteps}\leg{(v6)}{\angle}{\arabic{legSteps}}}}}
}

\newcommand{\dPentLegs}[7]{
\setcounter{legSteps}{0}
\def\zeroAngle{-90-45}\def\spread{45}\setcounter{offset}{-1}\addtocounter{offset}{#1}
\ifthenelse{#1=0}{}{\ifthenelse{#1=1}{\stepcounter{legSteps}\leg{(v1)}{\zeroAngle}{\arabic{legSteps}};}{
\foreach\n in {1,...,#1}{\def\eph{\arabic{offset}}\def\angle{\zeroAngle-2*\n*\spread/\eph+#1*\spread/\eph+\spread/\eph}\stepcounter{legSteps}\leg{(v1)}{\angle}{\arabic{legSteps}}}}}

\def\zeroAngle{90+45}\def\spread{45}\setcounter{offset}{-1}\addtocounter{offset}{#2}
\ifthenelse{#2=0}{}{\ifthenelse{#2=1}{\stepcounter{legSteps}\leg{(v2)}{\zeroAngle}{\arabic{legSteps}};}{
\foreach\n in {1,...,#2}{\def\eph{\arabic{offset}}\def\angle{\zeroAngle-2*\n*\spread/\eph+#2*\spread/\eph+\spread/\eph}\stepcounter{legSteps}\leg{(v2)}{\angle}{\arabic{legSteps}}}}}

\def\zeroAngle{90}\def\spread{25}\setcounter{offset}{-1}\addtocounter{offset}{#3}
\ifthenelse{#3=0}{}{\ifthenelse{#3=1}{\stepcounter{legSteps}\leg{(v3)}{\zeroAngle}{\arabic{legSteps}};}{
\foreach\n in {1,...,#3}{\def\eph{\arabic{offset}}\def\angle{\zeroAngle-2*\n*\spread/\eph+#3*\spread/\eph+\spread/\eph}\stepcounter{legSteps}\leg{(v3)}{\angle}{\arabic{legSteps}}}}}

\def\zeroAngle{45}\def\spread{45}\setcounter{offset}{-1}\addtocounter{offset}{#4}
\ifthenelse{#4=0}{}{\ifthenelse{#4=1}{\stepcounter{legSteps}\leg{(v4)}{\zeroAngle}{\arabic{legSteps}};}{
\foreach\n in {1,...,#4}{\def\eph{\arabic{offset}}\def\angle{\zeroAngle-2*\n*\spread/\eph+#4*\spread/\eph+\spread/\eph}\stepcounter{legSteps}\leg{(v4)}{\angle}{\arabic{legSteps}}}}}

\def\zeroAngle{-45}\def\spread{45}\setcounter{offset}{-1}\addtocounter{offset}{#5}
\ifthenelse{#5=0}{}{\ifthenelse{#5=1}{\stepcounter{legSteps}\leg{(v5)}{\zeroAngle}{\arabic{legSteps}};}{
\foreach\n in {1,...,#5}{\def\eph{\arabic{offset}}\def\angle{\zeroAngle-2*\n*\spread/\eph+#5*\spread/\eph+\spread/\eph}\stepcounter{legSteps}\leg{(v5)}{\angle}{\arabic{legSteps}}}}}

\def\zeroAngle{-90}\def\spread{25}\setcounter{offset}{-1}\addtocounter{offset}{#6}
\ifthenelse{#6=0}{}{\ifthenelse{#6=1}{\stepcounter{legSteps}\leg{(v6)}{\zeroAngle}{\arabic{legSteps}};}{
\foreach\n in {1,...,#6}{\def\eph{\arabic{offset}}\def\angle{\zeroAngle-2*\n*\spread/\eph+#6*\spread/\eph+\spread/\eph}\stepcounter{legSteps}\leg{(v6)}{\angle}{\arabic{legSteps}}}}}

\def\zeroAngle{180}\def\spread{20}\setcounter{offset}{-1}\addtocounter{offset}{#7}
\ifthenelse{#7=0}{}{\ifthenelse{#7=1}{\stepcounter{legSteps}\leg{(v7)}{\zeroAngle}{\arabic{legSteps}};}{
\foreach\n in {1,...,#7}{\def\eph{\arabic{offset}}\def\angle{\zeroAngle-2*\n*\spread/\eph+#7*\spread/\eph+\spread/\eph}\stepcounter{legSteps}\leg{(v7)}{\angle}{\arabic{legSteps}}}}}
}

\newcommand{\dBoxLegs}[6]{
\setcounter{legSteps}{0}
\def\zeroAngle{-90-45}\def\spread{45}\setcounter{offset}{-1}\addtocounter{offset}{#1}
\ifthenelse{#1=0}{}{\ifthenelse{#1=1}{\stepcounter{legSteps}\leg{(v1)}{\zeroAngle}{\arabic{legSteps}};}{
\foreach\n in {1,...,#1}{\def\eph{\arabic{offset}}\def\angle{\zeroAngle-2*\n*\spread/\eph+#1*\spread/\eph+\spread/\eph}\stepcounter{legSteps}\leg{(v1)}{\angle}{\arabic{legSteps}}}}}

\def\zeroAngle{90+45}\def\spread{45}\setcounter{offset}{-1}\addtocounter{offset}{#2}
\ifthenelse{#2=0}{}{\ifthenelse{#2=1}{\stepcounter{legSteps}\leg{(v2)}{\zeroAngle}{\arabic{legSteps}};}{
\foreach\n in {1,...,#2}{\def\eph{\arabic{offset}}\def\angle{\zeroAngle-2*\n*\spread/\eph+#2*\spread/\eph+\spread/\eph}\stepcounter{legSteps}\leg{(v2)}{\angle}{\arabic{legSteps}}}}}

\def\zeroAngle{90}\def\spread{25}\setcounter{offset}{-1}\addtocounter{offset}{#3}
\ifthenelse{#3=0}{}{\ifthenelse{#3=1}{\stepcounter{legSteps}\leg{(v3)}{\zeroAngle}{\arabic{legSteps}};}{
\foreach\n in {1,...,#3}{\def\eph{\arabic{offset}}\def\angle{\zeroAngle-2*\n*\spread/\eph+#3*\spread/\eph+\spread/\eph}\stepcounter{legSteps}\leg{(v3)}{\angle}{\arabic{legSteps}}}}}

\def\zeroAngle{45}\def\spread{45}\setcounter{offset}{-1}\addtocounter{offset}{#4}
\ifthenelse{#4=0}{}{\ifthenelse{#4=1}{\stepcounter{legSteps}\leg{(v4)}{\zeroAngle}{\arabic{legSteps}};}{
\foreach\n in {1,...,#4}{\def\eph{\arabic{offset}}\def\angle{\zeroAngle-2*\n*\spread/\eph+#4*\spread/\eph+\spread/\eph}\stepcounter{legSteps}\leg{(v4)}{\angle}{\arabic{legSteps}}}}}

\def\zeroAngle{-45}\def\spread{45}\setcounter{offset}{-1}\addtocounter{offset}{#5}
\ifthenelse{#5=0}{}{\ifthenelse{#5=1}{\stepcounter{legSteps}\leg{(v5)}{\zeroAngle}{\arabic{legSteps}};}{
\foreach\n in {1,...,#5}{\def\eph{\arabic{offset}}\def\angle{\zeroAngle-2*\n*\spread/\eph+#5*\spread/\eph+\spread/\eph}\stepcounter{legSteps}\leg{(v5)}{\angle}{\arabic{legSteps}}}}}

\def\zeroAngle{-90}\def\spread{25}\setcounter{offset}{-1}\addtocounter{offset}{#6}
\ifthenelse{#6=0}{}{\ifthenelse{#6=1}{\stepcounter{legSteps}\leg{(v6)}{\zeroAngle}{\arabic{legSteps}};}{
\foreach\n in {1,...,#6}{\def\eph{\arabic{offset}}\def\angle{\zeroAngle-2*\n*\spread/\eph+#6*\spread/\eph+\spread/\eph}\stepcounter{legSteps}\leg{(v6)}{\angle}{\arabic{legSteps}}}}}
}

\newcommand{\bTLegs}[5]{
\setcounter{legSteps}{0}
\def\zeroAngle{-90-45}\def\spread{45}\setcounter{offset}{-1}\addtocounter{offset}{#1}
\ifthenelse{#1=0}{}{\ifthenelse{#1=1}{\stepcounter{legSteps}\leg{(v1)}{\zeroAngle}{{\color{labelcolor}\arabic{legSteps}}};}{
\foreach\n in {1,...,#1}{\def\eph{\arabic{offset}}\def\angle{\zeroAngle-2*\n*\spread/\eph+#1*\spread/\eph+\spread/\eph}\stepcounter{legSteps}\leg{(v1)}{\angle}{{\color{labelcolor}\arabic{legSteps}}}}}}

\def\zeroAngle{90+45}\def\spread{45}\setcounter{offset}{-1}\addtocounter{offset}{#2}
\ifthenelse{#2=0}{}{\ifthenelse{#2=1}{\stepcounter{legSteps}\leg{(v2)}{\zeroAngle}{{\color{labelcolor}\arabic{legSteps}}};}{
\foreach\n in {1,...,#2}{\def\eph{\arabic{offset}}\def\angle{\zeroAngle-2*\n*\spread/\eph+#2*\spread/\eph+\spread/\eph}\stepcounter{legSteps}\leg{(v2)}{\angle}{{\color{labelcolor}\arabic{legSteps}}}}}}

\def\zeroAngle{70}\def\spread{25}\setcounter{offset}{-1}\addtocounter{offset}{#3}
\ifthenelse{#3=0}{}{\ifthenelse{#3=1}{\stepcounter{legSteps}\leg{(v3)}{\zeroAngle}{{\color{labelcolor}\arabic{legSteps}}};}{
\foreach\n in {1,...,#3}{\def\eph{\arabic{offset}}\def\angle{\zeroAngle-2*\n*\spread/\eph+#3*\spread/\eph+\spread/\eph}\stepcounter{legSteps}\leg{(v3)}{\angle}{{\color{labelcolor}\arabic{legSteps}}}}}}

\def\zeroAngle{0}\def\spread{45}\setcounter{offset}{-1}\addtocounter{offset}{#4}
\ifthenelse{#4=0}{}{\ifthenelse{#4=1}{\stepcounter{legSteps}\leg{(v4)}{\zeroAngle}{{\color{labelcolor}\arabic{legSteps}}};}{
\foreach\n in {1,...,#4}{\def\eph{\arabic{offset}}\def\angle{\zeroAngle-2*\n*\spread/\eph+#4*\spread/\eph+\spread/\eph}\stepcounter{legSteps}\leg{(v4)}{\angle}{{\color{labelcolor}\arabic{legSteps}}}}}}

\def\zeroAngle{-70}\def\spread{25}\setcounter{offset}{-1}\addtocounter{offset}{#5}
\ifthenelse{#5=0}{}{\ifthenelse{#5=1}{\stepcounter{legSteps}\leg{(v5)}{\zeroAngle}{{\color{labelcolor}\arabic{legSteps}}};}{
\foreach\n in {1,...,#5}{\def\eph{\arabic{offset}}\def\angle{\zeroAngle-2*\n*\spread/\eph+#5*\spread/\eph+\spread/\eph}\stepcounter{legSteps}\leg{(v5)}{\angle}{{\color{labelcolor}\arabic{legSteps}}}}}}
}

\newcommand{\dTLegs}[4]{
\setcounter{legSteps}{0}
\def\zeroAngle{180}\def\spread{45}\setcounter{offset}{-1}\addtocounter{offset}{#1}
\ifthenelse{#1=0}{}{\ifthenelse{#1=1}{\stepcounter{legSteps}\leg{(v1)}{\zeroAngle}{{\color{labelcolor}\arabic{legSteps}}};}{
\foreach\n in {1,...,#1}{\def\eph{\arabic{offset}}\def\angle{\zeroAngle-2*\n*\spread/\eph+#1*\spread/\eph+\spread/\eph}\stepcounter{legSteps}\leg{(v1)}{\angle}{{\color{labelcolor}\arabic{legSteps}}}}}}

\def\zeroAngle{90}\def\spread{55}\setcounter{offset}{-1}\addtocounter{offset}{#2}
\ifthenelse{#2=0}{}{\ifthenelse{#2=1}{\stepcounter{legSteps}\leg{(v2)}{\zeroAngle}{{\color{labelcolor}\arabic{legSteps}}};}{
\foreach\n in {1,...,#2}{\def\eph{\arabic{offset}}\def\angle{\zeroAngle-2*\n*\spread/\eph+#2*\spread/\eph+\spread/\eph}\stepcounter{legSteps}\leg{(v2)}{\angle}{{\color{labelcolor}\arabic{legSteps}}}}}}

\def\zeroAngle{0}\def\spread{60}\setcounter{offset}{-1}\addtocounter{offset}{#3}
\ifthenelse{#3=0}{}{\ifthenelse{#3=1}{\stepcounter{legSteps}\leg{(v3)}{\zeroAngle}{{\color{labelcolor}\arabic{legSteps}}};}{
\foreach\n in {1,...,#3}{\def\eph{\arabic{offset}}\def\angle{\zeroAngle-2*\n*\spread/\eph+#3*\spread/\eph+\spread/\eph}\stepcounter{legSteps}\leg{(v3)}{\angle}{{\color{labelcolor}\arabic{legSteps}}}}}}

\def\zeroAngle{-90}\def\spread{55}\setcounter{offset}{-1}\addtocounter{offset}{#4}
\ifthenelse{#4=0}{}{\ifthenelse{#4=1}{\stepcounter{legSteps}\leg{(v4)}{\zeroAngle}{{\color{labelcolor}\arabic{legSteps}}};}{
\foreach\n in {1,...,#4}{\def\eph{\arabic{offset}}\def\angle{\zeroAngle-2*\n*\spread/\eph+#4*\spread/\eph+\spread/\eph}\stepcounter{legSteps}\leg{(v4)}{\angle}{{\color{labelcolor}\arabic{legSteps}}}}}}
}

\newcommand{\tardiLegs}[5]{
\setcounter{legSteps}{0}
\def\zeroAngle{180}\def\spread{45}\setcounter{offset}{-1}\addtocounter{offset}{#1}
\ifthenelse{#1=0}{}{\ifthenelse{#1=1}{\stepcounter{legSteps}\leg{(v1)}{\zeroAngle}{{\color{labelcolor}\arabic{legSteps}}};}{
\foreach\n in {1,...,#1}{\def\eph{\arabic{offset}}\def\angle{\zeroAngle-2*\n*\spread/\eph+#1*\spread/\eph+\spread/\eph}\stepcounter{legSteps}\leg{(v1)}{\angle}{{\color{labelcolor}\arabic{legSteps}}}}}}

\def\zeroAngle{90}\def\spread{55}\setcounter{offset}{-1}\addtocounter{offset}{#2}
\ifthenelse{#2=0}{}{\ifthenelse{#2=1}{\stepcounter{legSteps}\leg{(v2)}{\zeroAngle}{{\color{labelcolor}\arabic{legSteps}}};}{
\foreach\n in {1,...,#2}{\def\eph{\arabic{offset}}\def\angle{\zeroAngle-2*\n*\spread/\eph+#2*\spread/\eph+\spread/\eph}\stepcounter{legSteps}\leg{(v2)}{\angle}{{\color{labelcolor}\arabic{legSteps}}}}}}

\def\zeroAngle{0}\def\spread{45}\setcounter{offset}{-1}\addtocounter{offset}{#3}
\ifthenelse{#3=0}{}{\ifthenelse{#3=1}{\stepcounter{legSteps}\leg{(v3)}{\zeroAngle}{{\color{labelcolor}\arabic{legSteps}}};}{
\foreach\n in {1,...,#3}{\def\eph{\arabic{offset}}\def\angle{\zeroAngle-2*\n*\spread/\eph+#3*\spread/\eph+\spread/\eph}\stepcounter{legSteps}\leg{(v3)}{\angle}{{\color{labelcolor}\arabic{legSteps}}}}}}

\def\zeroAngle{-90}\def\spread{55}\setcounter{offset}{-1}\addtocounter{offset}{#4}
\ifthenelse{#4=0}{}{\ifthenelse{#4=1}{\stepcounter{legSteps}\leg{(v4)}{\zeroAngle}{{\color{labelcolor}\arabic{legSteps}}};}{
\foreach\n in {1,...,#4}{\def\eph{\arabic{offset}}\def\angle{\zeroAngle-2*\n*\spread/\eph+#4*\spread/\eph+\spread/\eph}\stepcounter{legSteps}\leg{(v4)}{\angle}{{\color{labelcolor}\arabic{legSteps}}}}}}

\def\zeroAngle{180}\def\spread{35}\setcounter{offset}{-1}\addtocounter{offset}{#5}
\ifthenelse{#5=0}{}{\ifthenelse{#5=1}{\stepcounter{legSteps}\leg{(v5)}{\zeroAngle}{{\color{labelcolor}\arabic{legSteps}}};}{\ifthenelse{#5=2}{\stepcounter{legSteps}\leg{(v5)}{180}{{\color{labelcolor}\arabic{legSteps}}};\stepcounter{legSteps}\leg{(v5)}{0}{{\color{labelcolor}\arabic{legSteps}}};}{
\foreach\n in {1,...,#5}{\def\eph{\arabic{offset}}\def\angle{\zeroAngle-2*\n*\spread/\eph+#5*\spread/\eph+\spread/\eph}\stepcounter{legSteps}\leg{(v5)}{\angle}{{\color{labelcolor}\arabic{legSteps}}}}}}}

}

\newcommand{\contourVerts}[7]{
\ifthenelse{#1=1}{\node at (v1) [whiteDot] {};}{\ifthenelse{#1=2}{\node at (v1) [blueDot] {};}{\ifthenelse{#1=4}{\node at (v1) [blackDot] {};}{\ifthenelse{#1=3}{\node at (v1) [bdot] {};\node at (v1) [compositeDot] {};
}{\ifthenelse{#1=0}{\node at (v1) [bdot] {};}{}}}}}

\ifthenelse{#2=1}{\node at (v2) [whiteDot] {};}{\ifthenelse{#2=2}{\node at (v2) [blueDot] {};}{\ifthenelse{#2=4}{\node at (v2) [blackDot] {};}{\ifthenelse{#2=3}{\node at (v2) [bdot] {};\node at (v2) [compositeDot] {};
}{\ifthenelse{#2=0}{\node at (v2) [bdot] {};}{}}}}}

\ifthenelse{#3=1}{\node at (v3) [whiteDot] {};}{\ifthenelse{#3=2}{\node at (v3) [blueDot] {};}{\ifthenelse{#3=4}{\node at (v3) [blackDot] {};}{\ifthenelse{#3=3}{\node at (v3) [bdot] {};\node at (v3) [compositeDot] {};
}{\ifthenelse{#3=0}{\node at (v3) [bdot] {};}{}}}}}

\ifthenelse{#4=1}{\node at (v4) [whiteDot] {};}{\ifthenelse{#4=2}{\node at (v4) [blueDot] {};}{\ifthenelse{#4=4}{\node at (v4) [blackDot] {};}{\ifthenelse{#4=3}{\node at (v4) [bdot] {};\node at (v4) [compositeDot] {};
}{\ifthenelse{#4=0}{\node at (v4) [bdot] {};}{}}}}}

\ifthenelse{#5=1}{\node at (v5) [whiteDot] {};}{\ifthenelse{#5=2}{\node at (v5) [blueDot] {};}{\ifthenelse{#5=4}{\node at (v5) [blackDot] {};}{\ifthenelse{#5=3}{\node at (v5) [bdot] {};\node at (v5) [compositeDot] {};
}{\ifthenelse{#5=0}{\node at (v5) [bdot] {};}{}}}}}
\ifthenelse{#6=1}{\node at (v6) [whiteDot] {};}{\ifthenelse{#6=2}{\node at (v6) [blueDot] {};}{\ifthenelse{#6=4}{\node at (v6) [blackDot] {};}{\ifthenelse{#6=3}{\node at (v6) [bdot] {};\node at (v6) [compositeDot] {};
}{\ifthenelse{#6=0}{\node at (v6) [bdot] {};}{}}}}}
\ifthenelse{#7=1}{\node at (v7) [whiteDot] {};}{\ifthenelse{#7=2}{\node at (v7) [blueDot] {};}{\ifthenelse{#7=4}{\node at (v7) [blackDot] {};}{\ifthenelse{#7=3}{\node at (v7) [bdot] {};\node at (v7) [compositeDot] {};
}{\ifthenelse{#7=0}{\node at (v7) [bdot] {};}{}}}}}
}

\graphicspath{{figures/}}

\titleformat{\section}{\centering\normalsize\normalfont\bf}{\thesection}{0em}{}
\hypersetup{pdftitle={},pdfcreator={},linkcolor=[rgb]{0.15,0.35,0.75},colorlinks=true,citecolor=[rgb]{0.675,0,0.2},urlcolor=[rgb]{0.15,0.35,0.65}}
\newcommand{\fwbox}[2]{\text{\makebox[#1][c]{$\hspace{-150pt}\displaystyle#2\hspace{-150pt}$}}}
\newcommand{\fwboxL}[2]{\text{\makebox[#1][l]{$#2$}}}
\newcommand{\fwboxR}[2]{\text{\makebox[#1][r]{$#2$}}}
\newcommand{\bigger}[1]{\raisebox{-0.95pt}{\scalebox{1.25}{$#1$}}}

\newcommand{\eq}[1]{\vspace{-3.5pt}\begin{equation}\hspace{2pt}#1\hspace{-0pt}\vspace{-3.5pt}\end{equation}}

\newcommand{\fig}[3]{\raisebox{#1}{\includegraphics[scale=#2]{#3}}}

\newcommand{\equivR}{\fwbox{13.5pt}{\hspace{-0pt}\fwboxR{0pt}{\raisebox{0.47pt}{\hspace{1.45pt}:\hspace{-3pt}}}=\fwboxL{0pt}{}}}
\newcommand{\equivL}{\fwbox{13.5pt}{\fwboxR{0pt}{}=\fwboxL{0pt}{\raisebox{0.47pt}{\hspace{-3pt}:\hspace{1.45pt}}}}}

\newcommand{\br}[1]{\mathbin{\hspace{-1.5pt}\big[\hspace{-2.75pt}\big[#1\big]\hspace{-2.75pt}\big]\hspace{-1.5pt}}}

\definecolor{hblue}{rgb}{0,0,0.575}
\definecolor{hred}{rgb}{0.575,0.0,0.225}
\definecolor{hteal}{rgb}{0.0,0.545,0.7451}
\definecolor{perm}{rgb}{0.1,0.45,0.85}

\renewcommand{\phi}{\varphi}

\begin{document}
\title{\texorpdfstring{All Two-Loop, Color-Dressed Six-Point Amplitude Integrands in sYM}{All Two-Loop, Color-Dressed, Six-Point Amplitude Integrands in sYM}}
\author{Jacob~L.~Bourjaily}%\email{bourjaily@psu.edu}
\affiliation{Institute for Gravitation and the Cosmos, Department of Physics,\\Pennsylvania State University, University Park, PA 16802, USA}
\affiliation{Niels Bohr International Academy and Discovery Center, Niels Bohr Institute,\\University of Copenhagen, Blegdamsvej 17, DK-2100, Copenhagen \O, Denmark}
\author{Cameron~Langer}%\email{ckl5552@psu.edu}
\affiliation{Institute for Gravitation and the Cosmos, Department of Physics,\\Pennsylvania State University, University Park, PA 16802, USA}
\author{Yaqi~Zhang}%\email{yjz5289@psu.edu}
\affiliation{Institute for Gravitation and the Cosmos, Department of Physics,\\Pennsylvania State University, University Park, PA 16802, USA}

%%%%%%%%%%%%%%%%%%%%%%%%%%%%%%%%%%%%%%%%%%%%%%%%%%%%%%%%%%%%%%%%%%%%%%%%%%%%%%%%%%%%%%%%%%%

%%%%%%%%%%%%%%%%%%%%%%%%%%%%%%%%%%%%%%%%%%%%%%%%%%%%%%%%%%%%%%%%%%%%%%%%%%%%%%%%%%%%%%%%%%%
\begin{abstract}%
We construct all N${}^{k}$MHV six-particle amplitude integrands in color-dressed, maximally supersymmetric Yang-Mills theory at two loops in a single prescriptive basis of master integrands.
\end{abstract}
\maketitle
%%%%%%%%%%%%%%%%%%%%%%%%%%%%%%%%%%%%%%%%%%%%%%%%%%%%%%%%%%%%%%%%%%%%%%%%%%%%%%%%%%%%%%%%%%%

%%%%%%%%%%%%%%%%%%%%%%%%%%%%%%%%%%%%%%%%%%%%%%%%%%%%%%%%%%%%%%%%%%%%%%%%%%%%%%%%%%%%%%%%%%%
\vspace{-15pt}\section{Introduction}\label{introduction_section}\vspace{-14pt}
%%%%%%%%%%%%%%%%%%%%%%%%%%%%%%%%%%%%%%%%%%%%%%%%%%%%%%%%%%%%%%%%%%%%%%%%%%%%%%%%%%%%%%%%%%%
%
Much has been learned about the nature of perturbative scattering amplitudes in quantum field theory using unitarity-based methods to construct representations of loop \emph{integrands} in terms of some pre-chosen basis of Feynman-like integrands---specifically, integrals involving some graph of scalar Feynman propagators with particular loop-dependent numerators. At one loop and in the planar limit, such investigations have led to enormous advances in our understanding: to the discovery of tree-level recursion relations \cite{Britto:2004ap,Britto:2005fq} and their all-loop generalization \cite{ArkaniHamed:2010kv}; the discovery of dual-conformal (and, ultimately Yangian) symmetry of planar, maximally supersymmetric Yang-Mills theory \cite{Drummond:2006rz,Alday:2007hr,Drummond:2008vq,Drummond:2009fd}; to the connection between on-shell functions and Grassmannian integrals \cite{ArkaniHamed:2009sx,ArkaniHamed:2009dg,Kaplan:2009mh,Bourjaily:2010kw,Arkani-Hamed:2014bca}, positroid varieties \cite{ArkaniHamed:2012nw,Bourjaily:2012gy}, and the amplituhedron \cite{ArkaniHamed:2010gg,Arkani-Hamed:2013jha,Arkani-Hamed:2013kca}.

Much of this progress flowed out of the investigation of particular amplitudes---often at or just beyond the reach of previous methods. Even for processes with as few as six particles, specific calculations of observable-like quantities such as the ratio of two helicity amplitudes \cite{Bern:2008ap,Kosower:2010yk} or the finite remainders of exponentiated amplitudes \cite{DelDuca:2009au,DelDuca:2010zg,Bourjaily:2019jrk,Bourjaily:2019vby} led to powerful new tools and remarkable insights into the mathematical structure underlying quantum field theory (see e.g.~\cite{Goncharov:2010jf,Dixon:2011nj,Dixon:2011pw,Dixon:2013eka,Dixon:2014iba,Dixon:2014voa,Dixon:2014xca,Dixon:2015iva,Caron-Huot:2019vjl}). 

Even for the simplest quantum field theories, however, much less is understood beyond the planar limit. This is in part due to the complexities of color-dressing and in part due to the non-existence of any preferred `routing' of the loop momentum variables. In the planar limit, (symmetrized) dual-momentum coordinates defined by the planar dual-graphs of Feynman diagrams not only provide a natural and universal choice of loop variables, but also suggest a natural organization of loop-dependent numerators according to essentially na\"ive power-counting---the scaling of an integrand as any loop momentum is taken to infinity. For example, it is well-known that amplitude integrands in planar sYM are dual-conformally invariant, and the space of dual-conformal integrands provides a natural basis in which any multiplicity amplitude can be expressed (see e.g.~\cite{Drummond:2006rz,Alday:2007hr,Drummond:2008vq,Drummond:2009fd,ArkaniHamed:2010gh,Bourjaily:2013mma,Bourjaily:2015jna,Bourjaily:2017wjl}).

Beyond the planar limit, however, although there is considerable evidence that amplitudes in sYM have `good' ultraviolet behavior (see e.g.~\cite{Arkani-Hamed:2014via,Bourjaily:2018omh}), the precise meaning of this statement at the integrand-level remains to be fully clarified. Roughly speaking, all amplitudes in this theory are expected to be expressible in a basis of integrands which scale like a `box' or better at infinity; but defining this behavior for non-planar integrands is far from obvious. 

Recently, a graph-theoretic definition of power-counting suitable for non-planar loop integrands was described in \cite{Bourjaily:2020qca}. For any bounded multiplicity or fixed spacetime dimension, the authors of \cite{Bourjaily:2020qca} described how to define and enumerate a concrete set of loop integrands in which all amplitudes of any theory should be represented using the methods of unitarity. Roughly speaking, a non-planar integrand with $p$-gon power-counting is one which scales at infinite loop momentum like any one of a set of Feynman integrands defined as $p$-gon `scalars' (with loop independent numerators). (As defined by \cite{Bourjaily:2020qca}, a `scalar $p$-gon' is any Feynman graph with girth $p$ for which any edge-collapse would lower its girth.) For our purposes, it is worth noting that integrands with triangle power-counting include all those with box or better power-counting, and form a strict subset of those needed to represent amplitudes in the Standard Model; moreover, at two loops, they include \emph{all} integrands which integrate to functions with maximal transcendental weight in four dimensions, and therefore represent a space of master integrands in which all amplitudes in maximally supersymmetric ($\mathcal{N}\!=\!4$) Yang-Mills theory (`sYM') may be represented. 

The process of diagonalizing such a basis with respect to cuts (rendering it \emph{prescriptive} \cite{Bourjaily:2017wjl}) was recently described in detail in \cite{IllustratingBasisBuilding}, where we provided a complete, prescriptive basis of triangle power-counting integrands at two loops for six external particles. In this basis, all six-particle helicity amplitudes in color-dressed, sYM or maximal ($\mathcal{N}\!=\!8$) supergravity (`SUGRA') should be expressible. In this work, we use unitarity to determine the coefficients of all six-particle N$^k$MHV amplitudes in this basis---namely, the MHV, NMHV, and N$^2$MHV ($\overline{\text{MHV}}$) amplitudes, of which the NMHV amplitude is entirely new. These coefficients are each simple \emph{leading singularities}, expressible as (fully color-dressed) on-shell functions, valid for any choice of gauge group.

%%%%%%%%%%%%%%%%%%%%%%%%%%%%%%%%%%%%%%%%%%%%%%%%%%%%%%%%%%%%%%%%%%%%%%%%%%%%%%%%%%%%%%%%%%%
\vspace{-12pt}\section{Prescriptive Bases for Scattering Amplitudes}\vspace{-14pt}
%%%%%%%%%%%%%%%%%%%%%%%%%%%%%%%%%%%%%%%%%%%%%%%%%%%%%%%%%%%%%%%%%%%%%%%%%%%%%%%%%%%%%%%%%%%
%
The essential idea of generalized unitarity \cite{Bern:1994zx,Bern:1994cg,Britto:2004nc} is that any amplitude integrand $\mathcal{A}$ can be represented in a basis of pre-chosen, Feynman-like integrands $\{\mathcal{I}_J\}$,
\eq{\mathcal{A}=\sum_{J}\mathfrak{a}_J\,\mathcal{I}_J\,,\label{general_decomposition}}
with coefficients determined by \emph{cut-conditions} and given in terms of on-shell functions built from lower-loop (ultimately tree-level) scattering amplitudes. Any basis of loop integrands (viewed as differential forms on the space of internal loop momenta) is called \emph{prescriptive} if it is the \emph{cohomological-dual} of a spanning set of compact, maximum-dimensional integration contours \cite{Bourjaily:2017wjl,Bourjaily:2020hjv,Bourjaily:2021vyj,IllustratingBasisBuilding}. That is, a basis is prescriptive if there exists a set of cycles $\{\Omega_K\}$ such that the period matrix of $\{\mathcal{I}_J\}$,
\vspace{-5pt}\eq{\oint\limits_{\Omega_K}\mathcal{I}_J=\delta_{J}^{K}\,\label{prescriptivity_condition}}
is diagonal. This is called the \emph{prescriptivity condition}. Provided this is the case, the coefficients $\mathfrak{a}_i$ appearing in the representation of amplitudes (\ref{general_decomposition}) are simply individual \emph{leading singularities}---on-shell functions defined as
\eq{\mathfrak{a}_J\equivR\oint\limits_{\Omega_J}\mathcal{A}\,.\label{ls_coefficients}}

In the above discussion, the index `$J$' is a \emph{collective} index encoding all \emph{distinct} integrands or contours. We clarify the precise meaning of this summation in the following section.  

Importantly, nothing about this construction depends on how the internal loop momenta are parameterized as all these degrees of freedom are fully integrated-out in both (\ref{prescriptivity_condition}) and (\ref{ls_coefficients}). Thus, this story applies equally well to planar as well as non-planar amplitudes without reference to any \emph{particular} rational function of loop momenta to be called `the' loop integrand. 

Recently, we constructed a complete and prescriptive basis of two-loop, non-planar integrands for six external particles with triangle power-counting \cite{IllustratingBasisBuilding}. Because this basis \emph{contains} all integrands which integrate to functions with maximal transcendental weight, and all integrands with box (or better) power-counting, all helicity amplitudes in sYM should be expressible in this basis. Thus, to represent the six-particle MHV ($\simeq\!\overline{\text{MHV}}$) or NMHV amplitude in this basis, it suffices to compute all of the coefficients (\ref{ls_coefficients}) for either amplitude in the summand (\ref{general_decomposition}).

%%%%%%%%%%%%%%%%%%%%%%%%%%%%%%%%%%%%%%%%%%%%%%%%%%%%%%%%%%%%%%%%%%%%%%%%%%%%%%%%%%%%%%%%%%%
\vspace{-12pt}\section{Summing Terms for Amplitudes}\vspace{-14pt}
%%%%%%%%%%%%%%%%%%%%%%%%%%%%%%%%%%%%%%%%%%%%%%%%%%%%%%%%%%%%%%%%%%%%%%%%%%%%%%%%%%%%%%%%%%%
%
The basis of integrands constructed in \cite{IllustratingBasisBuilding} involved 87 integrand \emph{topologies} corresponding to specific Feynman-propagator graphs which encode their loop-dependent denominators; each of these topologies was then provided with a collection of specific triangle power-counting, loop-dependent numerators---the number of which is determined by the integrand's propagator-graph according to \cite{Bourjaily:2020qca}; their precise form of these numerators was determined by the requirement of prescriptivity (\ref{prescriptivity_condition}) with respect to a corresponding choice of contours. 

We may denote the integrand with the $i$th topology ($i\!\in\!\{1,\ldots,87\}$) and the $j$th loop-dependent numerator by $\mathcal{I}_{i}^j\equivR\mathcal{I}_i\mathfrak{n}_i^j$, where $\mathcal{I}_i$ consists of all loop-dependent denominators corresponding to some Feynman graph and $\mathfrak{n}_{i}^j$ denotes a particular, loop-dependent numerator for this integrand topology. In all, the basis described in  \cite{IllustratingBasisBuilding} consists of $373$ integrands, each written with a particular choice of external momenta flowing into the graph. 

Consider for example the integrand topology numbered $35$, with denominator encoded by the Feynman propagators given by
\vspace{-8pt}\eq{\fwboxR{0pt}{\mathcal{I}_{35}\;\;\bigger{\Leftrightarrow}\hspace{-7pt}}\npPbox{\draw[int](v1)--(v2)--(v3)--(v4)--(v5)--(v6)--(v3);\draw[int](v1)--(v5);\npPboxScalarEdges\npPboxLegs{1}{1}{0}{2}{0}{2}\intDots{6}}\fwboxL{0pt}{\hspace{-7pt}.}\label{eg_graph_35}\vspace{-5pt}}
For this integrand, there are 10 distinct loop-dependent numerators $\mathfrak{n}_{35}^j$. For example, the 6th such numerator is defined to be \cite{IllustratingBasisBuilding}
\eq{\mathfrak{n}_{35}^6\equivR{-}s_{12}\big(\br{a,e,d,c}{-}a^2d^2+c^2e^2\big)\,,\label{eg_num_35_6}}
in terms of the bracket $\br{\cdots}\equivR\mathrm{tr}_{+}(\cdots)$ and momenta flowing through the various edges of the graph (\ref{eg_graph_35}). This numerator has been fixed by the requirement that $\oint_{\Omega_{35}^6}\mathcal{I}_{35}^6=1$, where the contour $\Omega_{35}^6$ can be represented according to
\vspace{-8pt}\eq{\fwboxR{0pt}{\Omega_{35}^6\;\;\bigger{\Leftrightarrow}\hspace{-7pt}}\npPbox{\draw[dashed](v1)--(v2); \draw[int](v2)--(v3)--(v4)--(v5)--(v6)--(v3);\draw[int](v1)--(v5);\npPboxLegs{1}{1}{0}{2}{0}{2}\contourVerts{0}{0}{1}{4}{1}{4}{6}
}\fwboxL{0pt}{\hspace{-7pt}.}\label{eg_contour_35_6}\vspace{-5pt}}
In (\ref{eg_contour_35_6}), the particular choice of contour is encoded graphically as described in \cite{IllustratingBasisBuilding}---to which we refer the reader for complete details. (The apparent `contact-terms' appearing in (\ref{eg_num_35_6}) are fixed by prescriptivity (\ref{prescriptivity_condition}).)

The reader will notice that a particular choice of external legs have been used to encode the \emph{representative} basis integrand $\mathcal{I}_{i}^j$. Naturally, permutations of the external legs must be included as well in the summation (\ref{general_decomposition}). We may use `$\mathcal{I}_{i}^j[\sigma]$' to denote the integrand with topology $\mathcal{I}_i$ and numerator $\mathfrak{n}_{i}^j$ for which the external legs have been relabelled \emph{relative to the reference} integrand according to $[1,\ldots,6]\!\mapsto\![\sigma(1),\ldots,\sigma(6)]$ for any permutation $\sigma\!\in\!\mathfrak{S}_{6}$. We may use similar notation $\Omega_i^j[\sigma]$ to denote contours involving permuted leg labels relative to the seeds defined in \cite{IllustratingBasisBuilding} and $\mathfrak{a}_i^j[\sigma]$ for leading singularities. For example, 
\vspace{-8pt}\eq{\fwboxR{0pt}{\mathcal{I}_{35}[351264]\;\;\bigger{\Leftrightarrow}\hspace{-7pt}}\npPbox{\draw[int](v1)--(v2)--(v3)--(v4)--(v5)--(v6)--(v3);\draw[int](v1)--(v5);\npPboxScalarEdges\leg{(v1)}{-135}{3}\leg{(v2)}{135}{5}\leg{(v4)}{45}{1}\leg{(v4)}{-45}{2}\leg{(v6)}{205}{6}\leg{(v6)}{155}{4}\intDots{6}}\fwboxL{0pt}{\hspace{-7pt}.}}
Clearly, not all relabelings are inequivalent. For example, the integrand $\mathcal{I}_{35}^6$ has only $45\!=\!6!/(2^4)$ inequivalent permutations of the external legs, as permuting the pairs of legs $(3\!\leftrightarrow\!4)$ or $(5\!\leftrightarrow\!6)$; and swapping the sets $(3,4)\!\leftrightarrow\!(5,6)$ or flipping the graph vertically each leave the integrand invariant after relabeling the (arbitrary) edge labels.

In general, the number of inequivalent relabelings of a given integrand depends not only on the integrand topology (which may have nontrivial graph automorphisms), but also on the symmetries of its numerator---or, equivalently, the symmetries enjoyed by its defining contour of integration. Letting $\mathfrak{S}_{i}^j\!\subset\!\mathfrak{S}_6$ denote the subset of inequivalent relabelings of the external legs, the complete form of the summand in (\ref{general_decomposition}) can be written as 
\eq{\mathcal{A}=\sum_{i,j}\sum_{\sigma\in\mathfrak{S}_{i}^j}\mathfrak{a}_{i}^j[\sigma]\,\,\mathcal{I}_i^j[\sigma]\,.\label{detailed_summand}}

Actually, there is one final subtlety to mention. Although for any given distribution of external legs $\mathcal{I}_i^j[\sigma]$ encodes a distinct integrand for each $j$, it may be that the $j$th contour/integrand for one permutation of legs corresponds to the $k\!\neq\!j$th integrand/contour for a different permutation of legs. For example, consider two of the eight contours defining the integrands $\mathcal{I}_{10}^j$:
\vspace{-2pt}\eq{\begin{array}{@{}c@{}}\Omega_{10}^4[123456]\\[-10pt]\dPent{\dPentPlainEdges\dPentLegs{1}{1}{0}{1}{1}{0}{2}\contourVerts{1}{2}{1}{2}{1}{2}{4}}\\[-7pt]\text{(even)}\end{array}\begin{array}{@{}c@{}}\Omega_{10}^7[123456]\\[-10pt]\dPent{\dPentPlainEdges\dPentLegs{1}{1}{0}{1}{1}{0}{2}\contourVerts{2}{1}{2}{1}{2}{1}{4}}\\[-7pt]\text{(even)}\end{array}}
%\draw[dashed](v1)--(v2); \draw[int](v2)--(v3)--(v4)--(v5)--(v6)--(v3);\draw[int](v1)--(v5);\npPboxLegs{1}{1}{0}{2}{0}{2}\contourVerts{0}{0}{1}{4}{1}{4}{6}
%
It is easy to see that $\Omega_{10}^{4}[214356]\!\simeq\!\Omega_{10}^{7}[123456]$; as such, including both terms in the summand (\ref{detailed_summand}) would double-count such contributions. For all cases where this occurs, we choose to sum over all the leg-permuted images of only \emph{one} choice of the relevant seeds. 

Among the 373 contours used to define the basis given in \cite{IllustratingBasisBuilding}, a large fraction of these have manifestly no support for amplitudes in sYM. Specifically, the basis contours include 96 involving poles at infinity, 23 involving double-poles (associated with transcendental weight-drops), and 137 which have support on collinear \emph{but not soft} regions of loop momenta; all these integrands have vanishing leading singularities. In all, there are only 139 (of the 373) contours which define the basis that have support for amplitudes of some N${}^k$MHV-degree.

As most of the contours defining the integrand basis do not have support of amplitudes, relatively few are required to express each N$^k$MHV amplitude. In particular, we find that the MHV (or $\overline{\text{MHV}}\simeq$N${}^2$MHV) amplitude requires only 38 permutation-seeds, and the NMHV amplitude requires only 80. Summing over all relevant relabelings of external legs, these amplitudes involve a total of 7,680 and 21,135 terms, respectively. The required seed-terms and code to generate the complete summands are included in the ancillary files for this work.

%%%%%%%%%%%%%%%%%%%%%%%%%%%%%%%%%%%%%%%%%%%%%%%%%%%%%%%%%%%%%%%%%%%%%%%%%%%%%%%%%%%%%%%%%%%
\vspace{-12pt}\section{Six-Particle Leading Singularities at Two-Loops}\vspace{-14pt}
%%%%%%%%%%%%%%%%%%%%%%%%%%%%%%%%%%%%%%%%%%%%%%%%%%%%%%%%%%%%%%%%%%%%%%%%%%%%%%%%%%%%%%%%%%%
%
For either the MHV (or $\overline{\text{MHV}}$) or the NMHV amplitude, the leading singularities correspond to residues of amplitudes which involve some number of on-shell (and possibly soft) internal degrees of freedom. These are color-dressed on-shell functions which are fully Bose-symmetric in the legs entering each vertex tree-amplitude. The MHV-degree of a leading singularity is computed by the degrees of each of the amplitudes at its vertices. In particular, a leading singularity graph involving $n_I$ internal edges (after removing any soft edges) will have an overall N${}^k$MHV-degree of $k\!=\!\sum_v(k_v{+}2){-}(n_I{+}2)$---with MHV ($\overline{\text{MHV}}$) three-point vertices having degree $k_v\!=\!0,{-}1$, respectively. For any vertex in a cut-diagram involving some number $n_v\!>\!3$ three legs, $k_v$ can take any value $k_v\!\in\!\{0,\ldots,n_v{-}4\}$, allowing for some contours to support leading-singularities of multiple degrees. We encode the N${}^k$MHV-degree of a vertex amplitude by color, with $k_v={-}1,0,1$ represented by white, blue, and red, respectively (the only cases relevant for 6-particle amplitudes). 

To compute each color-dressed leading singularity relevant for an amplitude, we decompose each vertex amplitude into color- times kinematic-factors with precise ordering the momenta involved according to DDM \cite{DelDuca:1999rs}. Graphically, we can denote this decomposition by the following: for any subset of two legs labelled $\{{\color{hred}\alpha},{\color{hred}\beta}\}$, of each vertex amplitude of an on-shell diagram, we write
\vspace{-5pt}\eq{\scalebox{1.3}{\begin{tikzpicture}[scale=1,baseline=-3.05]\draw[optExt] (0,0)--(135:0.45);\draw[optExt] (0,0)--(45:0.45);\draw[rEdge] (0,0)--(-45:0.45);\draw[rEdge] (0,0)--(-135:0.45);\node[ephdot] at (105:0.275) [] {};\node[ephdot] at (75:0.275) [] {};\node at (-135:0.6) [] {{\footnotesize{\color{anchorLeg}$\alpha$}}};\node at (90:0.5) [] {{\footnotesize$A$}};\node at (-45:0.6) [] {{\footnotesize{\color{anchorLeg}$\beta$}}};\node[fullmhvBig] at (0,0) [] {};\end{tikzpicture}}\hspace{-5pt}=\sum_{{\color{hblue}\vec{a}}\in\mathfrak{S}\hspace{-1pt}(\hspace{-1pt}{\color{black}A}\hspace{-0.75pt})}\;\;\fig{-22.125pt}{1.25}{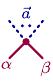}\hspace{-5pt}\times\hspace{-5pt}\fig{-22.125pt}{1.25}{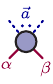}\,,\label{graphical_ddmification}\vspace{-5pt}}
where the `spherical' vertex denotes a Bose-symmetric, color-dressed tree amplitude of some N$^k$MHV-degree (indicated by its color) while the flat circle appearing on the right hand side represents the kinematic-part of an `(color-)ordered' amplitude involving a specific ordering of incoming states. In (\ref{graphical_ddmification}), the sum is taken over all $(n{-}2)!$ permutations ${\color{hblue}\vec{a}}\equivL({\color{hblue}a_1},\ldots,{\color{hblue}a_{{-}1}})$ of the \emph{unordered} set $A\equivR[n]\backslash\{{\color{hred}\alpha},{\color{hred}\beta}\}$ and with color-factors defined as 
\eq{\begin{split}f_{{\color{hred}\alpha\;\beta}}^{\;\,{\color{hblue}\vec{a}}}&\equivR\sum_{e_i}f^{{\color{hred}\alpha}\,{\color{hblue}a_1}\,e_1}f^{e_1\,{\color{hblue}a_2}\,e_2}\cdots f^{e_{\text{-}1}\,{\color{hblue}a_{\text{-}1}}\,{\color{hred}\beta}}\\&\equivL\fig{-15pt}{1}{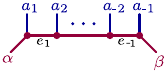}\equivL\fig{-15pt}{1}{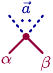}\,,\end{split}\label{notational_conventions_for_ddm_fs}}
with $f^{a\,b\,c}$ being the structure constants of the relevant gauge group. Applying this expansion at every vertex of a two-loop leading singularity results in an expansion in terms of kinematic-dependent on-shell functions built with amplitudes involving \emph{locally ordered} legs times general color-factors of the form 
\eq{\begin{split}\hspace{-30pt}f\!\bigger{\big[}{\color{hblue}\vec{a}},{\color{hblue}\vec{b}},{\color{hblue}\vec{c}}\bigger{\big]}&\equivR\fig{-17.125pt}{1}{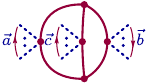}\\&=\sum_{{\color{hred}e_i}}f^{{\color{hred}e_1\,e_3\,e_6}}f_{{\color{hred}e_1\;e_2}}^{\;\;\,\,{\color{hblue}\vec{a}}}f_{{\color{hred}e_3\;e_4}}^{\;\;\,\,{\color{hblue}\vec{c}}}f_{{\color{hred}e_5\;e_6}}^{\;\;\,\,{\color{hblue}\vec{b}}}f^{{\color{hred}e_2\,e_5\,e_4}} .\end{split}\label{our_color_factors}}
This is similar to the decomposition described in \cite{Ochirov:2016ewn}. For examples of how this works for color-dressed leading singularities in sYM, see e.g.~\cite{Bourjaily:2019iqr,Bourjaily:2019gqu}. In the ancillary files for this work, code is given which can convert these color-factors into the color-trace basis for $\text{SU}(N_c)$ gauge-theory. 

To illustrate how this decomposition works, consider the case of contour $\Omega_{11}^1[123456]$, encoded graphically according to
\vspace{-5pt}\eq{\fwboxR{0pt}{\Omega_{11}^1[123456]\;\;\bigger{\Leftrightarrow}\hspace{-7pt}}\dPent{\dPentPlainEdges\dPentLegs{1}{1}{0}{1}{1}{1}{1}\contourVerts{1}{2}{1}{2}{1}{4}{2}}\fwboxL{0pt}{\hspace{-7pt}.}}
As the only N$^k$MHV degree allowed for the four-point vertex (involving leg `$5$') is $k_v\!=\!0$, this contour only has support for the NMHV amplitude. Using the DDM expansion described above, we would express $\mathfrak{a}_{11}^1[123456]$---the coefficient of $\mathcal{I}_{11}^1[123456]$---in (\ref{detailed_summand}) as
\vspace{-14pt}\eq{\begin{split}\hspace{-120pt}\fwbox{80pt}{\dPent{\dPentPlainEdges\dPentLegs{1}{1}{0}{1}{1}{1}{1}\lsVerts{1}{2}{1}{2}{1}{2}{2}}}=&\fwbox{80pt}{\dPent{\dPentPlainEdges\dPentLegs{1}{1}{0}{1}{1}{1}{1}\lsVertsOrd{1}{2}{1}{2}{1}{2}{2}}}\colorF{\coordinate (a1) at (210:0.7);\coordinate (a11) at (210:1);\coordinate(a2)at(150:0.7);\coordinate(a21)at(150:1);\coordinate (b1) at (45:0.7);\coordinate (b11) at (45:1);\coordinate(b2)at(0:0.7);\coordinate(b21)at(0:1);\coordinate(b3)at(-45:0.7);\coordinate(b31)at(-45:1);\coordinate(c1) at (0,0);\coordinate(c11)at(-0.3,0);\draw[fgraphEdge] (a1)--(a11);\draw[fgraphEdge] (a2)--(a21);\draw[fgraphEdge] (b1)--(b11);\draw[fgraphEdge] (b2)--(b21);\draw[fgraphEdge] (b3)--(b31);\draw[fgraphEdge] (c1)--(c11);\node at (a1) [fdot] {};\node at (a2) [fdot] {};\node at (b1) [fdot] {};\node at (b2) [fdot] {};\node at (b3) [fdot] {};\node at (c1) [fdot] {};\node[inner sep=1.5pt,anchor=east]at (a11) [] {\raisebox{-4pt}{{\footnotesize$\raisebox{-4pt}{1}$}}};
\node[inner sep=1.5pt,anchor=east]at (a21) [] {\raisebox{-4pt}{{\footnotesize$\raisebox{4pt}{2}$}}};
\node[inner sep=1.5pt,anchor=west]at (b11) [] {\raisebox{-4pt}{{\footnotesize$\raisebox{4pt}{3}$}}};
\node[inner sep=1.5pt,anchor=west]at (b21) [] {\raisebox{-4pt}{{\footnotesize$\raisebox{0pt}{4}$}}};
\node[inner sep=1.5pt,anchor=west]at (b31) [] {\raisebox{-4pt}{{\footnotesize$\raisebox{-4pt}{5}$}}};
\node[inner sep=1.5pt,anchor=east]at (c11) [] {\raisebox{-4pt}{{\footnotesize$\raisebox{-0pt}{6}$}}};}\hspace{-100pt}\\[-10pt]
&\fwboxL{0pt}{\!\!+}\fwbox{80pt}{\dPent{\dPentPlainEdges\leg{(v1)}{-135}{1}\leg{(v2)}{135}{2}\leg{(v4)}{45}{3}\leg{(v5)}{-45}{4}\leg{(v6)}{155}{5}\leg{(v7)}{180}{6}\lsVertsOrd{1}{2}{1}{2}{1}{2}{2}}}\colorF{\coordinate (a1) at (210:0.7);\coordinate (a11) at (210:1);\coordinate(a2)at(150:0.7);\coordinate(a21)at(150:1);\coordinate (b1) at (30:0.7);\coordinate (b11) at (30:1);\coordinate(b2)at(-30:0.7);\coordinate(b21)at(-30:1);\coordinate(c1) at (0,-0.2333);\coordinate(c11)at(-0.3,-0.2333);\coordinate(c2) at (0,0.2333);\coordinate(c21)at(-0.3,0.2333);\draw[fgraphEdge] (a1)--(a11);\draw[fgraphEdge] (a2)--(a21);\draw[fgraphEdge] (b1)--(b11);\draw[fgraphEdge] (b2)--(b21);\draw[fgraphEdge] (c2)--(c21);\draw[fgraphEdge] (c1)--(c11);\node at (a1) [fdot] {};\node at (a2) [fdot] {};\node at (b1) [fdot] {};\node at (b2) [fdot] {};\node at (c2) [fdot] {};\node at (c1) [fdot] {};\node[inner sep=1.5pt,anchor=east]at (a11) [] {\raisebox{-4pt}{{\footnotesize$\raisebox{-4pt}{1}$}}};
\node[inner sep=1.5pt,anchor=east]at (a21) [] {\raisebox{-4pt}{{\footnotesize$\raisebox{4pt}{2}$}}};
\node[inner sep=1.5pt,anchor=west]at (b11) [] {\raisebox{-4pt}{{\footnotesize$\raisebox{4pt}{3}$}}};
\node[inner sep=1.5pt,anchor=west]at (b21) [] {\raisebox{-4pt}{{\footnotesize$\raisebox{0pt}{4}$}}};
\node[inner sep=1.5pt,anchor=east]at (c21) [] {\raisebox{-4pt}{{\footnotesize$\raisebox{0pt}{6}$}}};
\node[inner sep=1.5pt,anchor=east]at (c11) [] {\raisebox{-4pt}{{\footnotesize$\raisebox{-0pt}{5}$}}};}\hspace{-10pt}\hspace{-100pt}\end{split}\label{ls_coeff_of_11_1}}
~\\[-20pt]
where the color-factors have been defined in (\ref{our_color_factors}) and the kinematic functions are on-shell functions involving specific-orderings of the legs involved at each vertex amplitude. These factors can be readily computed in terms of products of ordered amplitudes evaluated for the particular on-shell, internal momenta; but such functions were classified in \cite{Bourjaily:2016mnp}, and we have chosen to use their conventions here. Thus, we may identify the relevant kinematic-dependent pieces as being given by
\vspace{-6pt}\eq{\fwbox{0pt}{\begin{array}{@{}c@{}c@{}}
\dPent{\dPentPlainEdges\dPentLegs{1}{1}{0}{1}{1}{1}{1}\lsVertsOrd{1}{2}{1}{2}{1}{2}{2}}&\dPent{\dPentPlainEdges\leg{(v1)}{-135}{1}\leg{(v2)}{135}{2}\leg{(v4)}{45}{3}\leg{(v5)}{-45}{4}\leg{(v6)}{155}{5}\leg{(v7)}{180}{6}\lsVertsOrd{1}{2}{1}{2}{1}{2}{2}}\\[-2pt]
\fwboxR{0pt}{={-}}\mathfrak{f}_2^c[263451]&\fwboxR{0pt}{={-}}\mathfrak{f}_5[145632]\end{array}}}
(with signs dictated by the oriented contour of integration and the condition that $\oint_{\Omega_i^j}\mathcal{I}_i^j=1$). We refer the reader to \cite{Bourjaily:2016mnp} for details on the definitions of these 6-point NMHV on-shell functions; but we have included the definitions of those required for the 2-loop NMHV amplitude in the ancillary files for this work.

\vspace{-12pt}\subsubsection{The Six-Particle MHV Amplitude}\vspace{-14pt}
The relevant \emph{kinematic parts} of on-shell functions for MHV amplitudes (at any loop order or multiplicity) were classified in \cite{Arkani-Hamed:2014bca} and can be represented at two loops concretely in terms of 
a function
\eq{\fwbox{0pt}{\Gamma\big[{\color{hblue}(a_1,\ldots,a_{{-}1})},{\color{hblue}(b_1,\ldots,b_{{-}1})},{\color{hblue}(c_1,\ldots,c_{{-}1})}\big]}}
defined in \cite{Bourjaily:2019iqr,Bourjaily:2019gqu} (see also \cite{Bourjaily:2018omh}). This function is invariant under permutations of the ordering of its arguments and under cyclic rotations of each argument separately. 

All non-vanishing leading singularities of the six particle MHV amplitude were given in \cite{Bourjaily:2019iqr}; these correspond to 38 particular contours for (\ref{ls_coefficients}). As these 38 contours were among those used in the prescriptive basis of ref.~\cite{IllustratingBasisBuilding}, each of these coefficients are essentially the same (while the integrands are rather different). Beyond permuting the leg-labels for the permutation-seeds of each basis integrand of ref.~\cite{IllustratingBasisBuilding} (used here) relative to those used in ref.~\cite{Bourjaily:2019iqr}, these coefficients are identical. In our ancillary files, we have merely relabeled these permutation-seed coefficients (and modified the sum of inequivalent terms accordingly).

\vspace{-12pt}\subsubsection{The Six-Particle NMHV Amplitude}\vspace{-14pt}
For the six-particle NMHV amplitude, the kinematic part of all leading singularities (for arbitrary loop-order) were classified in ref.~\cite{Bourjaily:2016mnp}. For each of the defining contours of the triangle power-counting basis, identifying the correct kinematic function from among those classified in \cite{Bourjaily:2016mnp} was done by explicit calculation---by computing the products of on-shell \mbox{(cyclically-ordered, tree-)}amplitudes evaluated on the corresponding maximal-cut in loop-integrand space, and summing over all the states that could be exchanged between them. 

In all, we find 80 non-vanishing leading singularities for the contours $\{\Omega_i^j\}$ defining the basis in \cite{IllustratingBasisBuilding}. Explicit formulae for each---expressed in terms of the functions classified in \cite{Bourjaily:2016mnp}---are given in the ancillary files of this work. 

Interestingly, of the 10 classes of leading singularities enumerated in \cite{Bourjaily:2016mnp}, the only ones that are needed at two loops are $\mathfrak{f}_1$, $\mathfrak{f}_2$, $\mathfrak{f}_2^c$, $\mathfrak{f}_3$, $\mathfrak{f}_4$, $\mathfrak{f}_5$, and $\mathfrak{f}_8$---with various permutations of the external legs as arguments. Actually, for $\mathfrak{f}_8$, the authors of \cite{Bourjaily:2016mnp} considered only the sum of particular solutions to the cut equations: $\mathfrak{f}_8^{\text{even}}\equivR\mathfrak{f}_{8}^+{+}\mathfrak{f}_{8}^-$, where the superscript indicates the sign of the square root in the solution to the final quadratic cut-equation; for us, both $\mathfrak{f}_8^{\text{even}}$ and $\mathfrak{f}_8^{\text{odd}}\equivR\mathfrak{f}_{8}^+{-}\mathfrak{f}_{8}^-$ are required. The set of kinematic factors appearing are not all functionally independent; they satisfy the relations also classified in \cite{Bourjaily:2016mnp}.

%%%%%%%%%%%%%%%%%%%%%%%%%%%%%%%%%%%%%%%%%%%%%%%%%%%%%%%%%%%%%%%%%%%%%%%%%%%%%%%%%%%%%%%%%%%
\vspace{-16pt}\subsection{Integration, Infrared Structure, and Regularization}\vspace{-14pt}
%%%%%%%%%%%%%%%%%%%%%%%%%%%%%%%%%%%%%%%%%%%%%%%%%%%%%%%%%%%%%%%%%%%%%%%%%%%%%%%%%%%%%%%%%%%
%
The particular basis described in ref.~\cite{IllustratingBasisBuilding} has several features that make it promising for loop integration and for exposing critical information about IR-structure of amplitudes. In particular, it is empirically true that prescriptive integrands are often maximally-transcendental and \emph{pure} \cite{ArkaniHamed:2010gh,Drummond:2010cz} and thus satisfy canonical, nilpotent differential equations \cite{Henn:2013pwa,Broedel:2018qkq,Herrmann:2019upk}. This should make them comparatively easy to integrate---for example, according to the methods outlined in \cite{Kotikov:1990kg,Kotikov:2000vn,Remiddi:1997ny,Henn:2013pwa}; but perhaps also more directly as illustrated in \cite{Bourjaily:2018aeq,Bourjaily:2019jrk,Bourjaily:2019vby}. 

Another aspect of the basis which may prove valuable is that it was fully divided into infrared-finite and divergent subspaces. This was achieved by choosing as many contours $\{\Omega_J\}$ as possible in the diagonalization (\ref{prescriptivity_condition}) to encompass regions responsible for infrared divergences. Moreover, the coefficients of each IR-divergent integral are \emph{manifestly} given by lower-loop expressions; thus, the universal behavior of divergences should be manifested, making it easier to cancel them prior to loop integration. Ideally, we are optimistic that the ratio of different helicity amplitudes could be rendered \emph{locally finite} in the sense of \cite{Bourjaily:2021ewt} in this basis, but we must leave such questions to future work (see e.g.~\cite{Becher:2009cu,Catani:1998bh})

In particular, the MHV amplitude involves 17 and 21 integrands (seed terms for leg-permutation sums) which are manifestly infrared-finite and divergent, respectively;  the NMHV amplitude involves 36 finite integrands and 44 divergent integrands.\\[-6pt]

One final comment is in order regarding regularization. Because the basis of integrands in ref.~\cite{IllustratingBasisBuilding} was defined in strictly four spacetime dimensions and the coefficients $\mathfrak{a}_J$ were computed using four-dimensional contours, our integrands are not ensured to give the correct \emph{regulated} expressions if integrated using dimensional regularization: $\mathcal{O}(\epsilon)$ corrections to coefficients can lead to finite corrections to divergent integrals. Nevertheless, we strongly suspect that all regulator dependence will cancel for any finite observable. 

Although dimensional regularization is unquestionably the most familiar and most widely used regulator, it is important to note that infrared divergences can be regulated faithfully using massive propagators---by going to the Higgs branch of the theory as in \cite{Henn:2010bk,Henn:2010ir}. Importantly, $\mathcal{O}(m^2)$ corrections to integrand coefficients always lead to $\mathcal{O}(m^2)$ contributions to regulated expressions (never canceling a divergence as in $\epsilon/\epsilon$ in dimensional regularization); as such, our unregulated expressions will yield the correct, regulated results on the Higgs branch.

%%%%%%%%%%%%%%%%%%%%%%%%%%%%%%%%%%%%%%%%%%%%%%%%%%%%%%%%%%%%%%%%%%%%%%%%%%%%%%%%%%%%%%%%%%%
\vspace{-16pt}\subsection{Consistency Checks}\vspace{-14pt}
%%%%%%%%%%%%%%%%%%%%%%%%%%%%%%%%%%%%%%%%%%%%%%%%%%%%%%%%%%%%%%%%%%%%%%%%%%%%%%%%%%%%%%%%%%%
%
For both the MHV and NMHV amplitudes, the planar parts---the leading (in $1/N_c$) coefficient of $\text{tr}(123456)$ in the expansion of the color-factors---were compared directly against the known results \cite{ArkaniHamed:2010gh,Bourjaily:2015jna}. For the sake of the reader, the ancillary files include the loop-momentum routing for each of non-vanishing contributions in terms of a universal choice for dual-momentum coordinates. As these formulae are novel (and non-manifestly dual-conformal) representations of these amplitudes, this is a highly non-trivial check on the correctness of our result. 

Beyond the planar limit, it is worth noting that relatively few leading-singularities appear among the contours chosen for the basis. All \emph{other} leading singularities of the amplitude must get matched indirectly in the basis via global residue theorems. Such identities were used to fix all relative signs of terms appearing in these expressions. We have checked that all such residue theorems for both the MHV and NMHV amplitudes are satisfied by our expressions---ensuring that \emph{all} leading singularities of each amplitude are matched.

\vspace{-16pt}\section{Content of the Ancillary Files}\vspace{-14pt}
Complete expressions for both amplitudes are given as ancillary files to this work available from the abstract page on the \texttt{arXiv}. These files include: a plain-text data file including lists of all non-vanishing seed-terms for both the MHV and NMHV amplitudes expressions, complete definitions of all relevant leading singularity coefficients, and expressions for all integrands from \cite{IllustratingBasisBuilding} with non-vanishing coefficients; a \textsc{Mathematica} package containing tools to manipulate, display, and evaluate these expressions; and an illustration notebook that demonstrates the usage of these tools. 

%%%%%%%%%%%%%%%%%%%%%%%%%%%%%%%%%%%%%%%%%%%%%%%%%%%%%%%%%%%%%%%%%%%%%%%%%%%%%%%%%%%%%%%%%%%
\vspace{-16pt}\section{Conclusions and Discussion}\vspace{-14pt}
%%%%%%%%%%%%%%%%%%%%%%%%%%%%%%%%%%%%%%%%%%%%%%%%%%%%%%%%%%%%%%%%%%%%%%%%%%%%%%%%%%%%%%%%%%%
%
The integration of non-planar Feynman integrands involving six particles is at or arguably just beyond our present state of the art. Indeed, it was only recently that the first non-planar amplitudes (in maximally supersymmetric theories) were computed for five-particles \mbox{\cite{Chicherin:2018yne,Abreu:2018aqd,Abreu:2019rpt,Chicherin:2019xeg}}. In this work we have provided explicit representations of both the NMHV and MHV six particle amplitude in precisely the same prescriptive basis of integrands---engineered to simplify the work of loop integration and to maximally expose the infrared structure of each amplitude. We suspect that there exists some non-planar analogue of the IR-finite ratio function of planar theories, and the first non-trivial instance of such should appear for six particles. Once the integrated expressions are found for these integrands, we anticipate the discovery of simplifications as dramatic as for in the case of planar amplitudes (see e.g.~\cite{DelDuca:2009au,Goncharov:2010jf})---and we hope that these will lead to similarly powerful new insights for amplitudes beyond the planar limit.

%\nopagebreak

%\newpage
%%%%%%%%%%%%%%%%%%%%%%%%%%%%%%%%%%%%%%%%%%%%%%%%%%%%%%%%%%%%%%%%%%%%%%%%%%%%%%%%%%%%%%%%%%%
\vspace{-12pt}\section{Acknowledgments}\vspace{-15pt}
%%%%%%%%%%%%%%%%%%%%%%%%%%%%%%%%%%%%%%%%%%%%%%%%%%%%%%%%%%%%%%%%%%%%%%%%%%%%%%%%%%%%%%%%%%%
The authors gratefully acknowledge fruitful conversations with Enrico Herrmann and Jaroslav Trnka. This project has been supported by an ERC Starting Grant \mbox{(757978)}, a grant from the Villum Fonden \mbox{(15369)}, by a grant from the US Department of Energy \mbox{(DE-SC00019066)}.

%
%\vspace{-14pt}
%\bibliographystyle{physics}
%\bibliography{amplitude_refs}
%\end{document}
%

\vspace{-14pt}
\providecommand{\href}[2]{#2}\begingroup\raggedright\endgroup

\end{document}